\documentclass[a4paper,11pt]{article}
\usepackage{amsfonts}
\usepackage{cite}

\usepackage{graphicx}
\graphicspath{{Figures/}}
\usepackage{amsmath}
\usepackage{amssymb}
\usepackage{latexsym}
\usepackage[colorlinks]{hyperref}
\usepackage{color}
\usepackage{float}
\usepackage{cite}
\usepackage{makeidx}
\usepackage{colortbl}
\usepackage{csquotes}
\usepackage[squaren]{SIunits}

\usepackage[textfont={footnotesize,sf},labelfont={color=blue,bf,sf},labelsep=endash]{caption}
\usepackage[position=top,labelfont={color=blue,bf,sf}]{subfig}
\usepackage{bm}

\usepackage[title]{appendix}

\newcommand{\da}{\dagger}

\newcommand{\bra}{\begin{array}}
	\newcommand{\era}{\end{array}}
\newcommand{\beq}{\begin{equation}}
	\newcommand{\eeq}{\end{equation}}
\newcommand{\bqr}{\begin{eqnarray}}
	\newcommand{\eqr}{\end{eqnarray}}

\def\BC{\bb C}
\def\_\BC{\bbi C}

\def\no2 {{\textstyle{n\over 2}}}

\begin{document}
	\begin{titlepage}
		\setcounter{page}{1}
		\renewcommand{\thefootnote}{\fnsymbol{footnote}}

		\begin{flushright}
		\end{flushright}
		
		\vspace{5mm}
		\begin{center}
			
			{\Large \bf {Tuning Gap 
				in	Corrugated Graphene with Spin Dependence 
			}}

			\vspace{5mm}
			
			{\bf Jaouad El-hassouny}$^{a}$,
			{\bf Ahmed Jellal\footnote{\sf a.jellal@ucd.ac.ma}}$^{b,c}$
			and
			{\bf El Houssine Atmani}$^{a}$
			
			\vspace{5mm}
			
			{$^{a}$\em Laboratory of Condensed Matter Physics
				and Renewed Energy, FST Mohammedia\\
				Hassan II University, Casablanca, Morocco}
			
			{$^{b}$\em Laboratory of Theoretical Physics,  
				Faculty of Sciences, Choua\"ib Doukkali University},\\
			{\em PO Box 20, 24000 El Jadida, Morocco}

			{$^{c}$\em Canadian Quantum  Research Center,
				204-3002 32 Ave Vernon, \\ BC V1T 2L7,  Canada}

			\vspace{30mm}

			\begin{abstract}

We study transmission in a system  consisting of a curved graphene surface as an arc (ripple) of circle connected  to two flat graphene sheets on the left and right sides. We introduce a mass term in the curved part and study the effect of a generated band gap in spectrum  on transport properties for  spin-up/-down. The tunneling analysis allows us to find all transmission and reflections channels {in terms of} the band gap. This later acts by decreasing   the transmissions with spin-up/-down 	but increasing {those}
with spin opposite, which exhibit {different behaviors}.   
We find resonances appearing in reflection with the  same spin, thus backscattering with a spin-up/-down is not null in ripple.   
{ome spatial  shifts for the total conduction are observed}
in our model and the magnitudes of these shifts can be efficiently controlled by adjusting the band gap. This high order tunability of the tunneling effect can be used to design highly accurate {devices} based on graphene.
				
			 \vspace{3cm}
			

			
			\end{abstract}
		\end{center}
	\end{titlepage}

	
	
	\section{Introduction}
	
 Graphene is an hexagonal rearrangement of carbon atoms \cite{Novoselov1,Zhang1} and actually remains among the amazing two-dimensional systems   discovered recently  in material science.
This because it exhibits interesting properties ranging from a linear dispersion relation to Klein tunneling paradox
 \cite{katsnelson2012graphene, torres2014introduction}. 
In addition, 
the band structures in graphene are described by a low energy effective theory 
similar to the massless Dirac-Weyl fermions  \cite{dresselhaus1998physical, neto2009electronic}. 
Graphene possesses  an exceptionally high mobility of the charge carriers.
{On the other hand, graphene stimulated the researchers to look for  other two-dimensional  materials. As a consequence, 
%
%
 a great number  with
	intriguing properties have been reported,
	covering metals, semiconductors, and
	insulators \cite{Tan2017, Pei2018,Guo2019,Khan2020,Tyagi2020}.
	
}

 However, 
the inability to control such mobility 
is of paramount concern in nanoelectronics.
This is due to
the lack of band gap in its energy spectrum, 
which means that electric current in graphene cannot be completely
shut off. Such
characteristic makes graphene unsuitable for the development of many
electronic devices
and essentially reduces its applicability industrial and technological.		
{Then, it is necessary to open a finite gap in
	the energy dispersions at $ K $ point, which can be achieved  by various experimental mechanisms. Indeed, 
an energy gap can be opened by  deposing graphene sheet on a 
substrate. Indeed, a gap of
0.26 eV is induced in graphene by using silicon carbide (SiC) as substrate
\cite{7777}. Also a gap of the order of 30 meV is produced by considering
the hexagonal boron nitride (hBN) \cite{8888, 9999}. 
The doping with boron \cite{1010, 1111} or nitrogen \cite{1212}
atoms can allow for 
 opening and controlling an energy gap as well. As another alternative  method one may  use the strain engineering 
to realize an opened  gap in graphene
 \cite{1313,Low2011,Guinea2012}. 
%
}

In recent years it has become evident that the physical properties of graphene can be changed
by manipulating it in an external way to 
control its conductivity. 
Among the various mechanisms that may affect the carrier mobility, the diffusion that could be induced by ripple \cite{katsnelson2008electron} appears to be the most natural since graphene sheets are naturally corrugated due to stress. It was shown that
the amplitude and orientation of the unidirectional ripples can be controlled by a change in the components of an applied strain
\cite{baimova2012unidirectional}.
Several works discussed how to introduce ripples  in graphene sheets in a controlled manner, and how to use such ripples \cite{Duan2011, Amal2010, Wang2011, Guinea2009, Miranda2009, Bao2009, Fasolino2007}. 
Additionally, ripples can be created and controlled in suspended graphene, including heat treatment \cite{Bao2009} and placing graphene in a specially prepared substrate. This is because the curvature of the surface affects the $\pi$ orbitals which determine the electronic properties of graphene.
{Ripples 
 are distortions of the planar structure
of the graphene,}
leading to measured charge mobilities much lower than theoretically predicted 
\cite{Naumis2017, Boggild2017}.	
	
The fundamental objective of this article is to extend the analysis in \cite{pudlak2015cooperative,pudlak2020spin} to a case when the corrugated graphene is subjected to an external delta deviation in mass term, which generates a band gap in the energy spectrum. We will attempt to answer how the added offset could be used to create a high efficiency polarized spin current in a corrugated graphene system. Then in the first stage,  we derive the energy spectrum and use the  transfer matrix to	
analytically obtain the full transmission and reflection channels. We show that the creation of  band energy 
acts by decreasing the transmissions with spin-up/-down but
increasing with spin opposite, which exhibit {different behaviors}. We find
resonances appearing in reflection with the same spin, thus backscattering with a spin-up/-down is
not null in ripple. Consequently, 
starting from some critical values of the band gap it appears that
the spin filter get affected, which is resulted in reduction of the channels.
Furthermore,
we show that the total conductance get affected by the band  gap in contrary to
the  case of null gap \cite{pudlak2015cooperative,pudlak2020spin}.
Generally, we show that the presence of a band gap can be used as a key tool to
control the transport properties of our system.

	The present paper is organized as follow. In section 2, we formulate our problem
	and determine the solutions of energy spectrum in curved and flat regions  of graphene. In section 3, we apply the boundary conditions at two interfaces to generate a  transfer matrix allowing us to derive the transmission and reflection channels. Subsequently, we obtain the total conductance using four transmission channels indexed by spin-up/-down.  
	We numerically study our results  and provide different discussions as well as analysis in section 4. Finally, we conclude our work.

	\section{Model setting}
	
\begin{figure}[H]
		\begin{center}
			\includegraphics[width=0.55\linewidth]
			{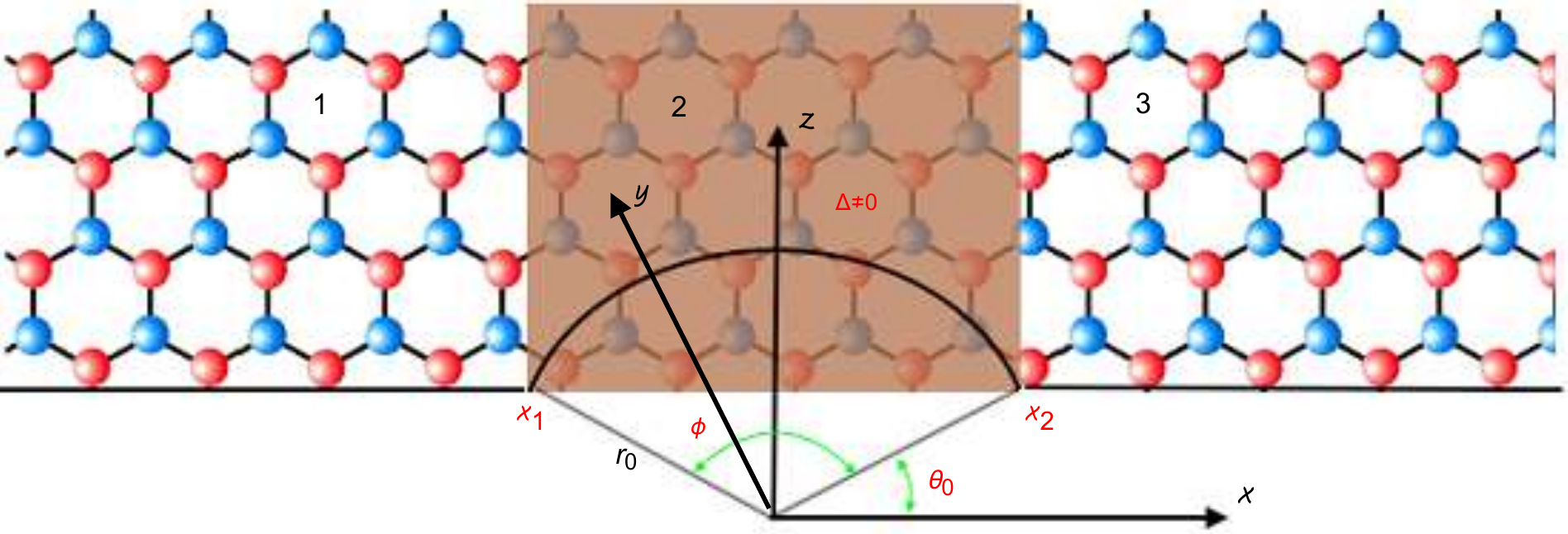}
		\end{center}
		\caption{Graphene system composed of there regions such that the medium  one is   an arc (ripple) of radius $r_{0}$ involving a mass term $\Delta$.}  
		\label{fig1}             
	\end{figure}

	We study  the scattering problem through 
	a  
	graphene involving in the central region a curved surface as arc  of  circle with radius $r_0$, or a ripple, and  mass term $\Delta $. In fact, we consider
	corrugated 
	graphene as depicted in Figure~\ref{fig1} with  region 1: $x\in]-\infty, x_1]$, region 3: $x\in[x_2, \infty[$ and region 2: $x\in [x_1, x_2]$, such that   $x_1=-x_2= -r_0\cos\theta_{0}$.
	In neglecting  edge effects we assume that  ${W} \gg {L},$ (${W}$ and ${L}$ being the width along $y$-  and  length along $x$-directions of graphene, respectively) and the ballistic electrons are injected to  ripple in the perpendicular direction, i.e. $k_{y}=0$. \\
	
	The present system can be described 
	in the basis $ (F_{A\uparrow}, 	F_{A\downarrow}, F_{B\uparrow},
	F_{B\downarrow}) $ of two sub-lattices $ (A,B) $
	by the Hamiltonian spin dependent
	\begin{equation}\label{eq1}
		H=
		\begin{pmatrix}
			0 & {D} \\
			{D}^{\dagger} & 0
		\end{pmatrix}
	\end{equation}
	such that the operators are given by
	\begin{eqnarray}
		&&	{D}=\gamma\left({p}_{x}-i {p}_{y}\right) \mathbb{I}+i \frac{\delta \gamma'}{4 r_0} {\sigma_{r}}-\frac{2 \delta \gamma p}{r_0} {\sigma_{y}}+ \Delta{\sigma_{\theta}}\\
		&&
		{D}^\da=\gamma\left({p}_{x}+i {p}_{y}\right) \mathbb{I}-i \frac{\delta \gamma'}{4 r_0} {\sigma_{r}}-\frac{2 \delta \gamma p}{r_0} {\sigma_{y}}+ \Delta{\sigma_{\theta}}
	\end{eqnarray}	
	where 
	${p}_{x}=- \frac{i\hbar}{r_0}{\partial}_{\theta}, {p}_{y}=-i\hbar {\partial}_{y}$,  ${\sigma_{r}}={\sigma_{x}} \cos \theta-{\sigma_{z}} \sin \theta$,
	${\sigma_{\theta}}=-{\sigma_{x}} \sin \theta-{\sigma_{z}} \cos \theta$ and 
	${\sigma}_{x, y, z}$ are  Pauli matrices.
	The set of involved parameters is given by  \cite{ando2000spin}
	\begin{align}
	\gamma=-\frac{\sqrt{3}}{2} V_{p p}^{\pi} a =\gamma_{0} a, \qquad 	
	\gamma^{\prime}=\frac{\sqrt{3}}{2}\left(V_{p p}^{\sigma}-V_{p p}^{\pi}\right) a=\gamma_{1} a, \qquad p=\frac{1-3 \gamma^{\prime}} {8\gamma} 
	\end{align}
	with $V_{p p}^{\alpha}$ and $V_{p p}^{\pi}$ are the transfer integrals for $\sigma$ and $\pi$ orbitals, respectively.  In the flat graphene, $a=$ $\sqrt{3} d \simeq 2.46 \AA$ is the length of the primitive translation vector, with $d$ is the distance between atoms in the unit cell. As for
	the intrinsic source of the spin-orbit coupling $\delta$
	we have 
	\begin{align}
	\delta=i \frac{ \hbar}{4 m_{e}^{2} c^{2}\epsilon_{\pi \sigma}}\left\langle x\left|\frac{\partial V}{\partial x} {p}_{y}-\frac{\partial V}{\partial y} {p}_{x}\right| y\right\rangle		
	\end{align}
with  the atomic potential $V$ and  $\epsilon_{\pi \sigma}=\epsilon_{2 p}^{\pi}-\epsilon_{2 p}^{\sigma}$, such that $\epsilon_{2 p}^{\sigma}$ is the energy of $\sigma$ orbitals (localized between carbon atoms)
and   $\epsilon_{2 p}^{\pi}$ is the energy of $\pi$ orbitals (directed perpendicular to the curved surface). In the next, for our  numerical purpose 
	it is convenient to  
	choose $\gamma=6.39\ \angstrom\ \electronvolt, \gamma^{\prime}=17.04\ \angstrom\ \electronvolt, \delta=0.01$,  $p=0.1$.

	To simply diagonalize {the Hamiltonian} $ H $ \eqref{eq1} we get rid of the $\theta$-dependence by making use of  the unitarty transformation in terms of $ \sigma_{y} $, which is  
	\begin{equation}
		{U}=
		\begin{pmatrix}
			e^{i \frac{\theta}{2} {\sigma}_{y}} & 0 \\
			0 & e^{i \frac{\theta}{2} {\sigma}_{y}}
		\end{pmatrix}
	\end{equation}
	and {then it transforms}    \eqref{eq1}  into the following 
	\begin{equation}
		\mathcal{H}={U} {H} {U}^{-1}
	\end{equation}
	giving rise to the Hamiltonian
	\begin{equation}\label{eqq8}
		\mathcal{H}
		= 
		\begin{pmatrix}
			0 & 0 & -\gamma\partial_{y}-\Delta-\mathrm{i} \frac{\gamma}{r_{0}} \partial_{\theta} & \mathrm{i}\left(\lambda_{y}+\lambda_{x}\right) \\
			0 & 0 & \mathrm{i}\left(\lambda_{y}-\lambda_{x}\right) &-\gamma\partial_{y}+\Delta-\mathrm{i} \frac{\gamma}{r_{0}} \partial_{\theta} \\
			\gamma\partial_{y}-\Delta-\mathrm{i} \frac{\gamma}{r_{0}} \partial_{\theta} & -\mathrm{i}\left(\lambda_{y}-\lambda_{x}\right) & 0 & 0 \\
			-\mathrm{i}\left(\lambda_{y}+\lambda_{x}\right) & \gamma\partial_{y}+\Delta-\mathrm{i} \frac{\gamma}{r_{0}} \partial_{\theta} & 0 & 0
		\end{pmatrix}
	\end{equation}
	where we have set  $\hbar=1$.   
	One can define  the total angular momentum by 
	\begin{equation}\label{eq7}
		{J}_{y}
		= 
		\begin{pmatrix}
			-\mathrm{i} \partial_{\theta}& \frac{	-\mathrm{i}}{2}& 0& 0  \\
			\frac{\mathrm{i}}{2}& -\mathrm{i}\partial_{\theta} & 0& 0 \\
			0& 0 & -\mathrm{i} \partial_{\theta}& \frac{	-\mathrm{i}}{2}  \\
			0& 0 &	\frac{\mathrm{i}}{2}& -\mathrm{i}\partial_{\theta}  
		\end{pmatrix}
	\end{equation}
	which transforms as $  \mathcal{J}_{y}={U} {J}_{y} {U}^{-1} $ and
	\begin{equation}\label{eq8}
		\mathcal{J}_{y}
		= 
		\begin{pmatrix}
			-\mathrm{i} \partial_{\theta}& 0& 0& 0  \\
			0& -\mathrm{i}\partial_{\theta} & 0& 0 \\
			0& 0 & -\mathrm{i} \partial_{\theta}& 0 \\
			0& 0 &	0& -\mathrm{i}\partial_{\theta} 
		\end{pmatrix}.
	\end{equation}

	Now based on the fact that 
	$ \left[\mathcal{H}, \mathcal{J}_{y}\right]=0 $ holds, we can use the separability of     eigenspinors  and then write $  \mathcal{F}(\theta, y) $ as
	\begin{equation}\label{eqq13}
		\mathcal{F}(\theta, y)=
		e^{\mathrm{i} m \theta} e^{\mathrm{i} k_{y} y}
		\begin{pmatrix}
			A \\
			B\\
			C\\
			D
		\end{pmatrix}
	\end{equation}
	where $ e^{\mathrm{i} m \theta} $ are eigensates of $ \mathcal{J}_y $ associated to the eigenvalues $ m=\pm \frac{1}{2}, \pm \frac{3}{2}, \cdots$. By injecting 
	\eqref{eqq8} and \eqref{eqq13} into the eigenvalue equation  $\mathcal{H} \mathcal{F}(\theta, y) =E \mathcal{F}(\theta, y)$ we find
	\begin{equation}\label{eq9}
		\begin{pmatrix}
			0 & 0 & t_{m}-\Delta-\mathrm{i} t_{y} & \mathrm{i}\left(\lambda_{y}+\lambda_{x}\right) \\
			0 & 0 & \mathrm{i}\left(\lambda_{y}-\lambda_{x}\right) & t_{m}+\Delta-\mathrm{i} t_{y} \\
			t_{m}-\Delta+\mathrm{i} t_{y} & -\mathrm{i}\left(\lambda_{y}-\lambda_{x}\right) & 0 & 0 \\
			-\mathrm{i}\left(\lambda_{y}+\lambda_{x}\right) & t_{m}+\Delta+\mathrm{i} t_{y} & 0 & 0
		\end{pmatrix}
		\begin{pmatrix}
			A \\
			B\\
			C\\
			D
		\end{pmatrix}
		=E \begin{pmatrix}
			A \\
			B\\
			C\\
			D
		\end{pmatrix}
	\end{equation}
{which allows}
	 to end up with four bands $ E_{s}^{s'} $ for the Hamiltonian $ H $
	\begin{equation}\label{eqq17}
		E_{s}^{s'}=s'\sqrt{\Delta^{2}+t_{m}^{2}+t_{y}^{2}+\lambda_{x}^{2}+\lambda_{y}^{2}+2s \sqrt{t_{m}^{2}\left(\Delta^{2}+\lambda_{x}^{2}\right)+\left(\Delta^{2}+t_{y}^{2}+\lambda_{x}^{2}\right) \lambda_{y}^{2}}}
	\end{equation}
	where  $ s,s'=\pm $ and we have set 
	$
	t_{m}=\frac{\gamma}{r_0} m$, $t_{y}=\gamma k_{y}$, $\lambda_{x}=\frac{\gamma}{2 r_0}(1+4 \delta p)$,  $\lambda_{y}=\frac{\delta \gamma^{\prime}}{4 r_0}
	$.
	Note that for normal incidence, i.e. $k_{y}=0,$ \eqref{eqq17} reduces to
	\begin{equation}\label{eqq18}
		E_{s}^{s'}=s'\sqrt{t_{m}^{2}+\lambda_{y}^{2}}+s\sqrt{\Delta^{2}+\lambda_{x}^{2}}
	\end{equation}
	which can be used to index the eigenvalues of $ \mathcal{J}_y $
	as 
	\begin{equation}\label{eqq24}
		m_s^{s'}=s'\frac{r_{0}}{\gamma} \sqrt{\left(E_{s}^{s'}-s \sqrt{\Delta^2+\lambda_{x}^2} \right)^{2}-\lambda_{y}^{2}} 
	\end{equation}
	a result that will be employed in the boundary conditions to determine
	the transmission and reflections coefficients.

\begin{figure}[H]
		\begin{center}
			\includegraphics[scale=0.43]{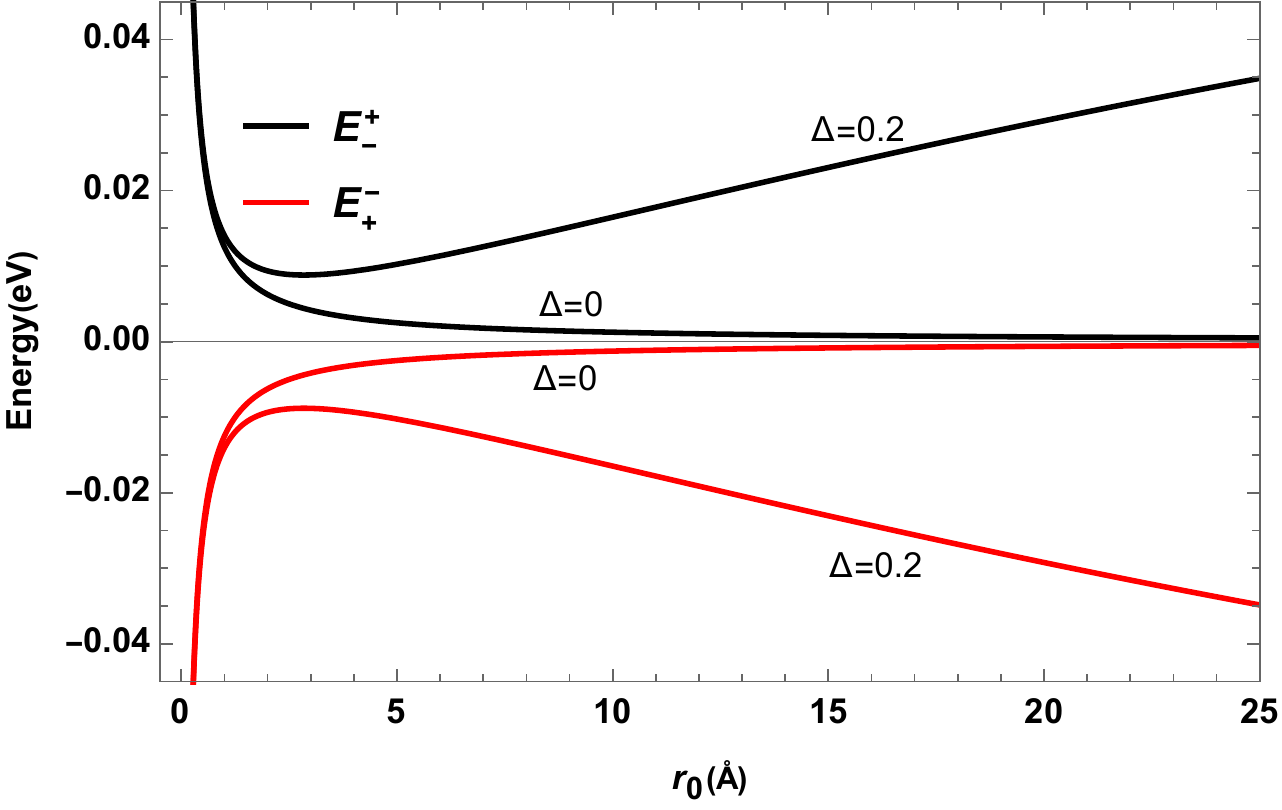} 
			\includegraphics[scale=0.43]{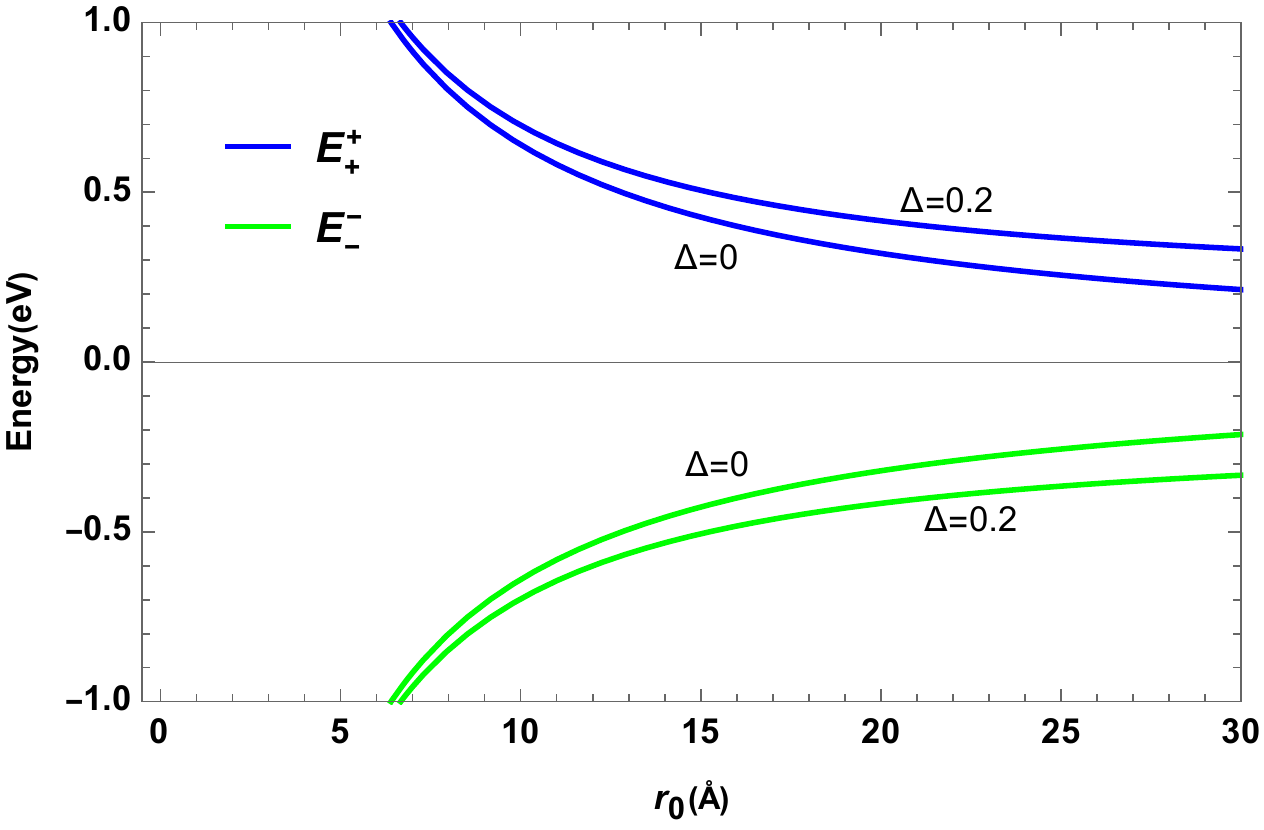} 
			\includegraphics[scale=0.43]{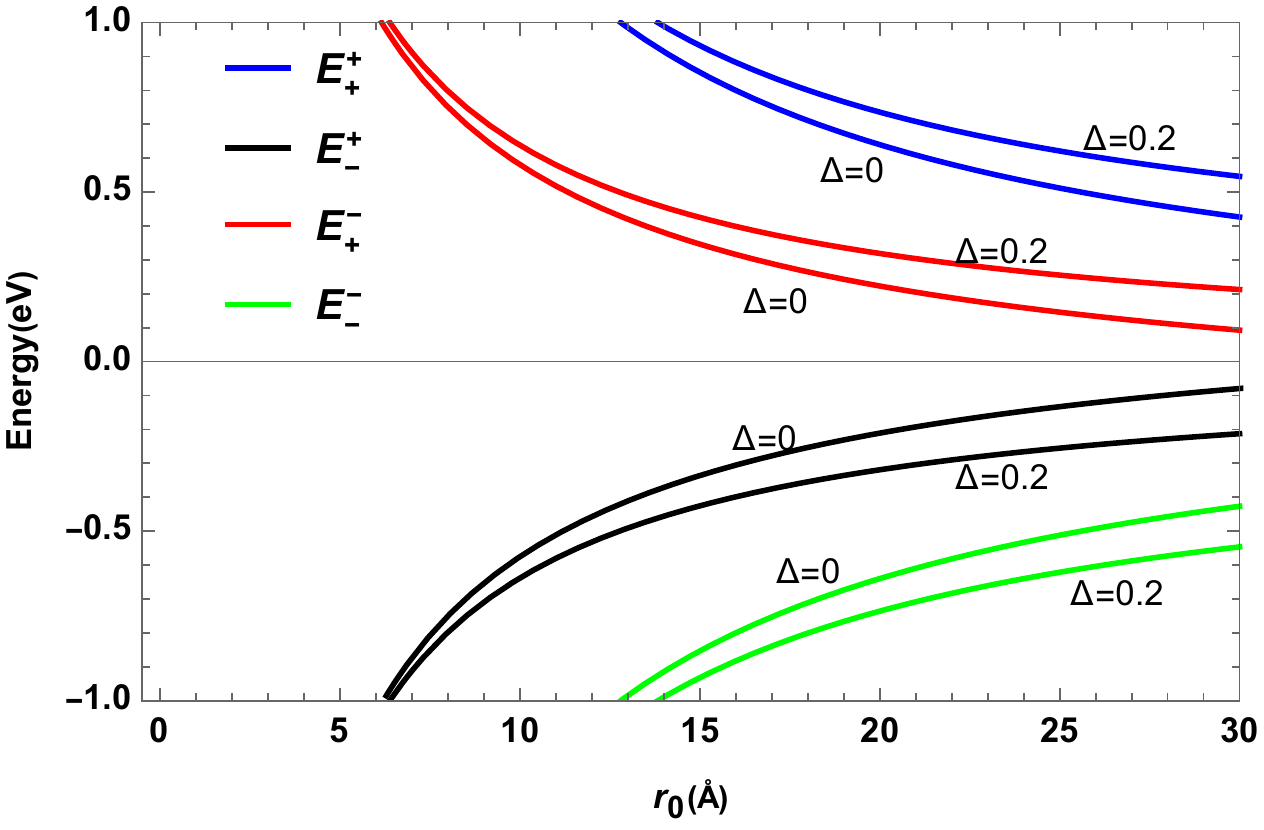}
		\end{center}
		\caption{(color online) The four bands  as a function of the radius $r_{0}$ for $ \delta=0.01,  p=0.1$ with $m=\frac{1}{2}$ (left  and middle panels), $m=\frac{3}{2}$ (right panel).  $E_{+}^{+}$  (blue), $E_{-}^{+}$ (black), $E_{+}^{-}$ (red), $E_{-}^{-}$ (green).  }
		\label{fig33}             
	\end{figure}

	To show the effect of the mass term $\Delta$, we plot the four bands  \eqref{eqq18} as a function of the radius of the curvature $r_{0}$ in Figure \ref{fig33}
	for $ \delta=0.01,  p=0.1$. For $m=\frac{1}{2}$ in left panel, we observe that at $r_{0} = 0$, the energies $E_{+}^{-}$ and $E_{-}^{+}$ have an infinite value. At $r_{0} \neq 0$, the difference between $E_{+}^{-}$ and $E_{-}^{+}$ is very small for $\Delta = 0$ but this difference increases with $\Delta = 0.2$. In the middle panel, at $\Delta = 0$, the difference between the energy $E_{+}^{+}$ and $E_{-}^{-}$ is greater than that of the left panel. On the other hand, at $\Delta = 0.2$, this difference increases.
	For $m=\frac{3}{2}$ in right panel, the shape of the curve of the four energies changes in the neighborhood of the interval $[0, 5]$ for the radius $r_{0}$. At $\Delta = 0.2$, the difference between $E_{+}^{-}$ and $E_{-}^{+}$ is greater than at $\Delta = 0$, as well as for $E_{+}^{+}$ and $E_{-}^{-}$. This shows that the introduction of the gap $\Delta$ is very necessary to control the differences between energies.\\

	To completely determine the solutions of energy spectrum, we solve  \eqref{eq9}
	to  end up with 
	the quantities
	\begin{eqnarray}
		&&A_{s,s'}(m)=s'\frac{\lambda_{x} \sqrt{t_{m}^{2}+\lambda_{y}^{2}}+s \lambda_{y} \sqrt{\Delta^{2}+\lambda_{x}^{2}}}{\lambda_{x} t_{m}-\lambda_{y} \Delta}\\
		&&
		B_{s,s'}(m)=- i s'\frac{\Delta \sqrt{t_{m}^{2}+\lambda_{y}^{2}}+s t_{m} \sqrt{\Delta^{2}+\lambda_{x}^{2}}}{\lambda_{x} t_{m}-\lambda_{y} \Delta}
		\\
		&&	C=1\\
		&& D_{s}(m)=- i\frac{t_{m}\Delta +\lambda_{y}\lambda_{x}+s\sqrt{(t_{m}^{2}+\lambda_{y}^{2})(\Delta^{2}+\lambda_{x}^{2})}}{\lambda_{x} t_{m}-\lambda_{y} \Delta}
	\end{eqnarray}
	and according to \eqref{eqq24} we have $m=m_s^{s'}$.
	Consequently the eigenspinors associated to four bands \eqref{eqq18} take the form
	{\begin{equation}
			\Psi_{2}^{s,s'}(\theta, m)=
			\begin{pmatrix}
				\cos\frac{\theta}{2}A_{s,s'}(m) -	\sin\frac{\theta}{2}	B_{s,s'}(m)\\
				\sin\frac{\theta}{2}A_{s,s'}(m) +	\cos\frac{\theta}{2}	B_{s,s'}(m)\\
				\cos\frac{\theta}{2}-	\sin\frac{\theta}{2}D_{s}(m)\\
				\sin\frac{\theta}{2}+	\cos\frac{\theta}{2}D_{s}(m)
			\end{pmatrix}
			e^{i m \theta}.
		\end{equation}
		As a result, in region 2 the wave function can be written as a superposition of all solutions for a curved surface in the form
		\begin{equation}
			\Psi_{2}(\theta)=	a_{+} \Psi_{2}^{+}\left(\theta, m_{+}^+\right)
			+b_{+} \Psi_{2}^{+}\left(\theta, m_{+}^-\right)+a_{-} \Psi_{2}^{-}\left(\theta, m_{-}^+\right)
			+b_{-} \Psi_{2}^{-}\left(\theta,m_{-}^-\right)
		\end{equation}
		$a_{\pm}$ and $ b_{\pm},$ denote the coefficients of the linear combination.
	}

	As for
	flat graphene (refers to regions 1 and 3), one can solve the eigenvalue equation to  derive the
	two band energy at $k_y=0$
	\begin{equation}
		E=\pm \gamma|k|
	\end{equation}
	and due to the energy conservation we have the relation $ E=E^{s'}_s $. The
	associated  eigenspinors are given by
	\begin{equation}
		\Psi_{1(3)}^{\tau}(x, k)=\frac{1}{2}
		\begin{pmatrix}
			\text{sign}(k) \\
			\text{sign}(k) \	i\tau \\
			1 \\
			i	\tau 
		\end{pmatrix}
		e^{i k  x}
	\end{equation}
	where  $ \tau=\pm$, $ k=k_{x} $ and    $\text{sign}(k)=\pm$ refers to  conductance and valence bands, respectively. As a result
	the eigenspinors  
	can be written a superposition of all possible solutions for flat graphene, such as
	in region 1
	\begin{equation}
		\Psi_{1}(x, k)=\alpha\Psi_{1}^{+}(x, k)+\beta\Psi_{1}^{-}(x, k)+r_{\uparrow}^{\xi} \Psi_{1}^{+}(x,-k)+r_{\downarrow}^{\xi} \Psi_{1}^{-}(x,-k)
	\end{equation}
	and in region 3
	\begin{equation}
		\Psi_{3}(x, k)=t_{\uparrow}^{\xi} \Psi_{3}^{+}(x, k)
		+t_{\downarrow}^{\xi} \Psi_{3}^{-}(x, k)
	\end{equation}
	with $ \alpha=0, \beta=1$ for spin-down  (${\xi}={\downarrow}$) and $ \alpha=1, \beta=0$ for spin-up (${\xi}={\uparrow}$) polarizations.  The coefficients $r_{\uparrow}^{\xi}, r_{\downarrow}^{\xi}, t_{\uparrow}^{\xi}, t_{\downarrow}^{\xi} $ denote eight channels of reflection and transmission.
	In the next, we will implement the above results to study some features of the present system. More precisely, we analyze the tunneling effect and discuss the influence of energy gap on transmission and reflection channels.

	\section{Transport properties}
	
	To determine the transmission and reflection amplitudes, we consider the continuity of eigenspinors at the two interfaces $(x_{1}=-r_0 \cos \theta_{0}, \theta_1=\pi+\theta_{0})$ and $(x_{2}=r_0 \cos \theta_{0}, \theta_2=2\pi-\theta_{0})$.   This process yields the set of equations
	\begin{eqnarray}
		&&
		\alpha\Psi_{1}^{+}(x_{1}, k)+\beta\Psi_{1}^{-}(x_{1}, k)+r_{\uparrow}^{\xi} \Psi_{1}^{+}(x_{1},-k)+r_{\downarrow}^{\xi} \Psi_{1}^{-}(x_{1},-k)=\nonumber \\ 
		&&a_{+} \Psi_{2}^{+}\left(\theta_1, m_{+}^+\right)
		+b_{+} \Psi_{2}^{+}\left(\theta_1, m_{+}^-\right)+a_{-} \Psi_{2}^{-}\left(\theta_1, m_{-}^+\right)
		+b_{-} \Psi_{2}^{-}\left(\theta_1,m_{-}^-\right)\label{eq19}
		\\
		&&
		\label{eq20}
		a_{+} \Psi_{2}^{+}\left(\theta_2, m_{+}^+\right)
		+b_{+} \Psi_{2}^{+}\left(\theta_2,m_{+}^-\right)+a_{-} \Psi_{2}^{-}\left(\theta_2, m_{-}^+\right)
		+b_{-} \Psi_{2}^{-}\left(\theta_2,m_{-}^-\right) =\nonumber \\
		&&t_{\uparrow}^{\xi} \Psi_{1}^{+}(x_{2}, k)
		+t_{\downarrow}^{\xi} \Psi_{1}^{-}(x_{2}, k)
	\end{eqnarray}
	where $  \xi=\downarrow, \uparrow$ refers to spin-up/-down. Now
	by eliminating the parameters $a_{\pm}$ and $ b_{\pm},$ we derive the relation
	\begin{equation}\label{eq21}
		\begin{pmatrix}
			\alpha \\
			\beta\\
			r_{\uparrow}^{\xi} \\
			r_{\downarrow}^{\xi}
		\end{pmatrix}
		=M_{0}^{-1} M\left(\frac{3\pi-\phi }{2}\right) {D}(-\phi) M^{-1}\left(\frac{3\pi+\phi }{2}\right) M_{0}
		\begin{pmatrix}
			t_{\uparrow}^{\xi} \\
			t_{\downarrow}^{\xi} \\
			0 \\
			0
		\end{pmatrix}
	\end{equation}
	and we have {introduced} the angle $\phi =\pi-2\theta_{0}.$
	Note that  for moving electrons  in \eqref{eq21} the configuration 
	$ \alpha=0, \beta=1$ is spin down polarization (${\xi}={\downarrow}$), while $ \alpha=1, \beta=0$ for spin up one (${\xi}={\uparrow}$). The involved matrix are given by
	\begin{equation}
		M_{0}=\begin{pmatrix}
			1 & 1 & -1 & -1 \\
			i  & -i & -i & i \\
			1 & 1 & 1 & 1 \\
			i & -i & i & -i
		\end{pmatrix},\qquad
		{D}(\phi)=\begin{pmatrix}
			e^{i m_{+} \phi} & 0 & 0 & 0 \\
			0 & e^{i m_{-} \phi} & 0 & 0 \\
			0 & 0 & e^{-i m_{+} \phi} & 0 \\
			0 & 0 & 0 & e^{-i m_{-} \phi}
		\end{pmatrix}
	\end{equation}
	and we have
	\begin{equation}
		M\left(\frac{3\pi+\phi }{2}\right)=\Lambda M_{A}
	\end{equation}
	such that $\Lambda$ is
	\begin{equation}
		\Lambda=\begin{pmatrix}
			\cos \frac{3\pi+\phi }{4} & -\sin  \frac{3\pi+\phi }{4} & 0 & 0 \\
			\sin \frac{3\pi+\phi }{4} & \cos  \frac{3\pi+\phi }{4} & 0 & 0 \\
			0 & 0 & \cos \frac{3\pi+\phi }{4} & -\sin  \frac{3\pi+\phi }{4} \\
			0 & 0 & \sin  \frac{3\pi+\phi }{4} & \cos  \frac{3\pi+\phi }{4}
		\end{pmatrix}
	\end{equation}
	as well as $M_{A}$ reads as
	\begin{equation}\label{eq26}
		M_{A}=\begin{pmatrix}
			A_{+,s'}(m_{+}) & A_{-,s'}(m_{-}) & A_{+,s'}(-m_{+}) & A_{-s'}(-m_{-}) \\
			B_{+,s'}(m_{+}) & B_{-,s'}(m_{-}) & B_{+,s'}(-m_{+}) & B_{-,s'}(-m_{-}) \\
			1 & 1 & 1 & 1 \\
			D_{+}(m_{+}) & D_{-}(m_{-}) & D_{+}(-m_{+}) & D_{-}(-m_{-}) \\
		\end{pmatrix}.
	\end{equation}
	{where}  $s'=+1$ for $|E| \geq\sqrt{\lambda_{x}^{2}+\Delta^2}$ and $s'=-1 $ for $|E|<\sqrt{\lambda_{x}^{2}+\Delta^2}$. Then from the above analysis
	we deduce the transfer matrix 
	\begin{equation}
		\mathcal{M}=M_{0}^{-1} \left(\Lambda M_{A}\right) 
		{D}(-\phi) \left(\Lambda M_{A}\right)^{-1}
		M_{0}=
		\left(
		\mathcal{M}_{ij}
		\right)
	\end{equation}
	with  $ i,j=1,2,3,4 $. After a lengthy and straightforward algebra, we end up with
	the transmission and reflection amplitudes for spin-up 
	\begin{eqnarray}
		&&t_{\uparrow}^{\uparrow}=\frac{\mathcal{M}_{22}}{\mathcal{M}_{22}\mathcal{M}_{11}-
			\mathcal{M}_{21}\mathcal{M}_{12}},
		\qquad
		t_{\downarrow}^{\uparrow}=-\frac{\mathcal{M}_{21}}{\mathcal{M}_{22}\mathcal{M}_{11}-
			\mathcal{M}_{21}\mathcal{M}_{12}} \label{322}
		\\
		&&
		r_{\uparrow}^{\uparrow}=\mathcal{M}_{31}t_{\uparrow}^{\uparrow}+\mathcal{M}_{32}
		t_{\downarrow}^{\uparrow},
		\qquad \qquad\ \
		r_{\downarrow}^{\uparrow}=\mathcal{M}_{41}t_{\uparrow}^{\uparrow}+\mathcal{M}_{42}t_{\downarrow}^{\uparrow}
	\end{eqnarray}
	as well as for spin-down 
	\begin{eqnarray}
		&&t_{\downarrow}^{\downarrow}=\frac{\mathcal{M}_{11}}{\mathcal{M}_{22}\mathcal{M}_{11}-\mathcal{M}_{21}\mathcal{M}_{12}},
		\qquad
		t_{\uparrow}^{\downarrow}=-\frac{\mathcal{M}_{12}}{\mathcal{M}_{22}\mathcal{M}_{11}
			-\mathcal{M}_{21}\mathcal{M}_{12}} \label{344}
		\\
		&&
		r_{\uparrow}^{\downarrow}=\mathcal{M}_{31} t_{\uparrow}^{\downarrow}+\mathcal{M}_{32} t_{\downarrow}^{\downarrow},
		\qquad\qquad \ \
		r_{\downarrow}^{\downarrow}=\mathcal{M}_{41}t_{\uparrow}^{\downarrow}+\mathcal{M}_{42}t_{\downarrow}^{\downarrow}.
	\end{eqnarray}
	From \eqref{322} and \eqref{344} we can derive the relation
	\begin{equation}
		t_{\uparrow}^{\uparrow}=\frac{\mathcal{M}_{22}}{\mathcal{M}_{21}} 
		t_{\downarrow}^{\uparrow}=
		\frac{\mathcal{M}_{22}}{\mathcal{M}_{11}} 
		t_{\downarrow}^{\downarrow}= \frac{\mathcal{M}_{22}}{\mathcal{M}_{12}} 
		t^{\downarrow}_{\uparrow}.
	\end{equation}
	Since the wave vector of input is the same as of output, then
	the associated  
	transmission and reflection 
	probabilities take the forms
	\begin{eqnarray}
		&& T_{\uparrow}^{\uparrow}=	|t_{\uparrow}^{\uparrow}|^{2},\qquad \ T_{\downarrow}^{\downarrow}= |t_{\downarrow}^{\downarrow}|^{2},\qquad
		T_{\uparrow}^{\downarrow}= |t_{\uparrow}^{\downarrow}|^{2}, \qquad
		T_{\downarrow}^{\uparrow}= |t_{\downarrow}^{\uparrow}|^{2}\\
		&&	R_{\uparrow}^{\uparrow}=	|r_{\uparrow}^{\uparrow}|^{2},\qquad R_{\downarrow}^{\downarrow}= |r_{\downarrow}^{\downarrow}|^{2},\qquad
		R_{\uparrow}^{\downarrow}= |r_{\uparrow}^{\downarrow}|^{2}, \qquad
		R_{\downarrow}^{\uparrow}= |r_{\downarrow}^{\uparrow}|^{2}.
	\end{eqnarray}

	The  conductance {associated to our system} can be obtained via the  Landauer-B{\"u}ttiker formula \cite{buttiker1985generalized,landauer1957,bundesmann2013spin} at zero temperature. 
	{It was shown that the conductance of one propagating channel can easily  be 
		generalized to any number of incoming and outgoing
		channels \cite{Pastawski2001,Pastawski2002,Pastawski2014}. 
		There are different conductances resulted from 
		the transmittances between states of definite momentum
		and spin projection at the contacts \cite{Imry1999}.
Consequently, 	the total conductance in
the linear response regime is given by summing over all the transmission channels
		}
%
%
	\begin{equation}
		G=\frac{e^{2}}{h}\left(T_{\uparrow}^{\uparrow}+T_{\downarrow}^{\downarrow}+T_{\uparrow}^{\downarrow}+T_{\downarrow}^{\uparrow}\right)
		=\frac{e^{2}}{h}
		\sum_{p, q=\xi} T_{p}^{q}.
	\end{equation}
	These results will  numerically be analyzed under suitable choices of the physics parameters involved in our system. This study will allow us to show the role that could play  the energy gap to modify the electronic properties of our system. 
	
	\section{Numerical results}
	
\begin{figure}[H]
		\begin{center}
			\subfloat[]{
				\includegraphics[scale=0.42]{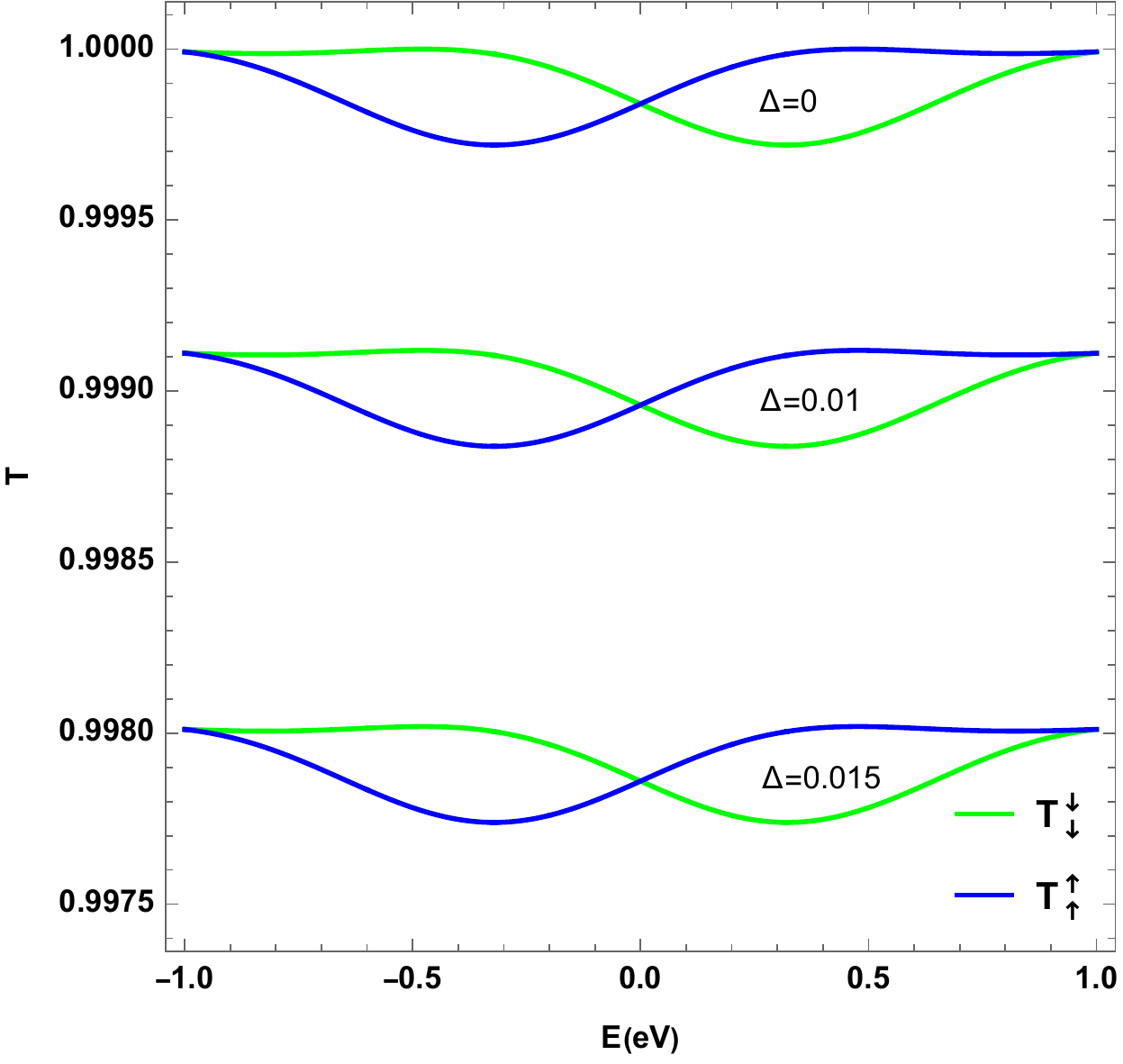}
				\label{subfigureaa}}
			\subfloat[]{
				\includegraphics[scale=0.42]{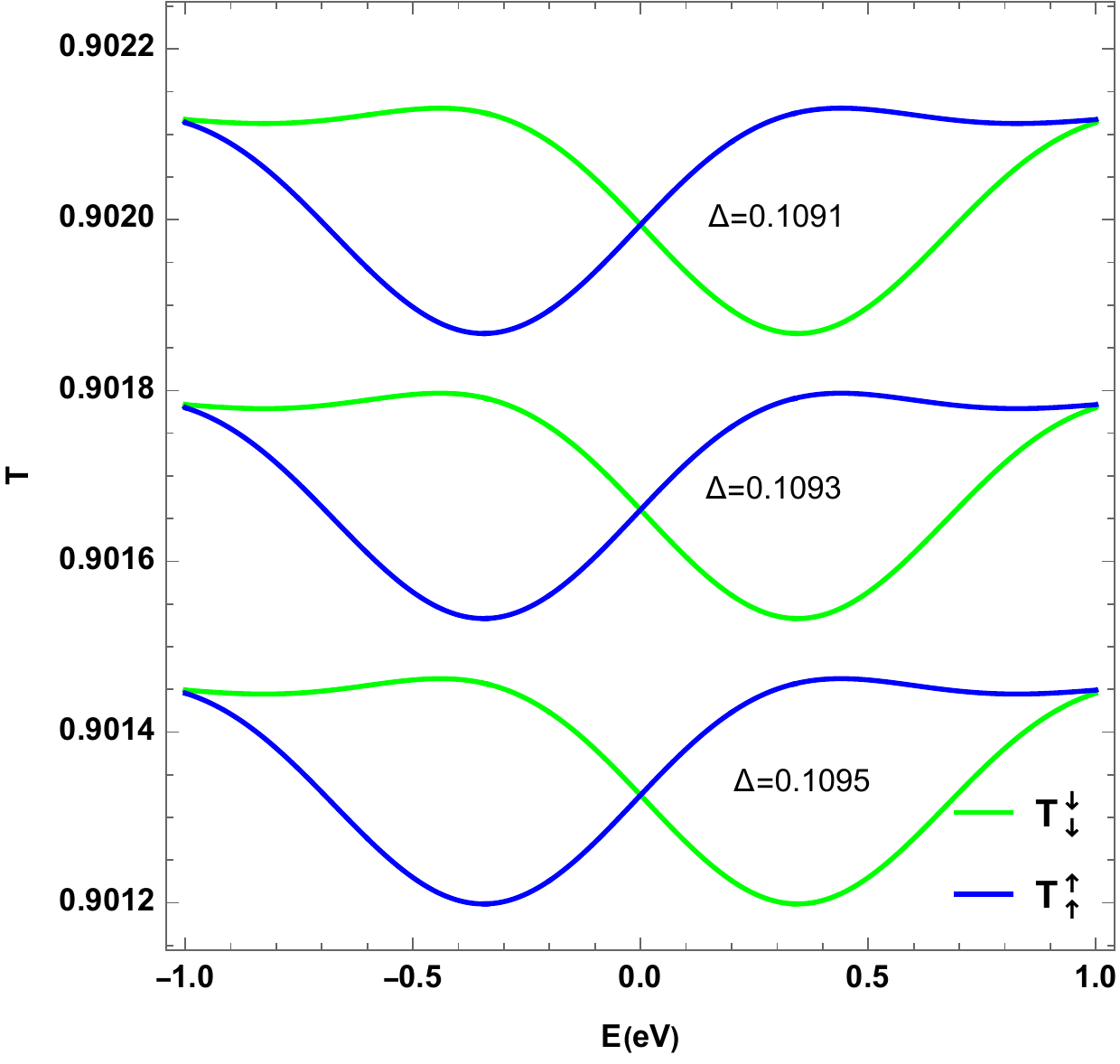}
				\label{subfigurebb}}
			\subfloat[]{
				\includegraphics[scale=0.42]{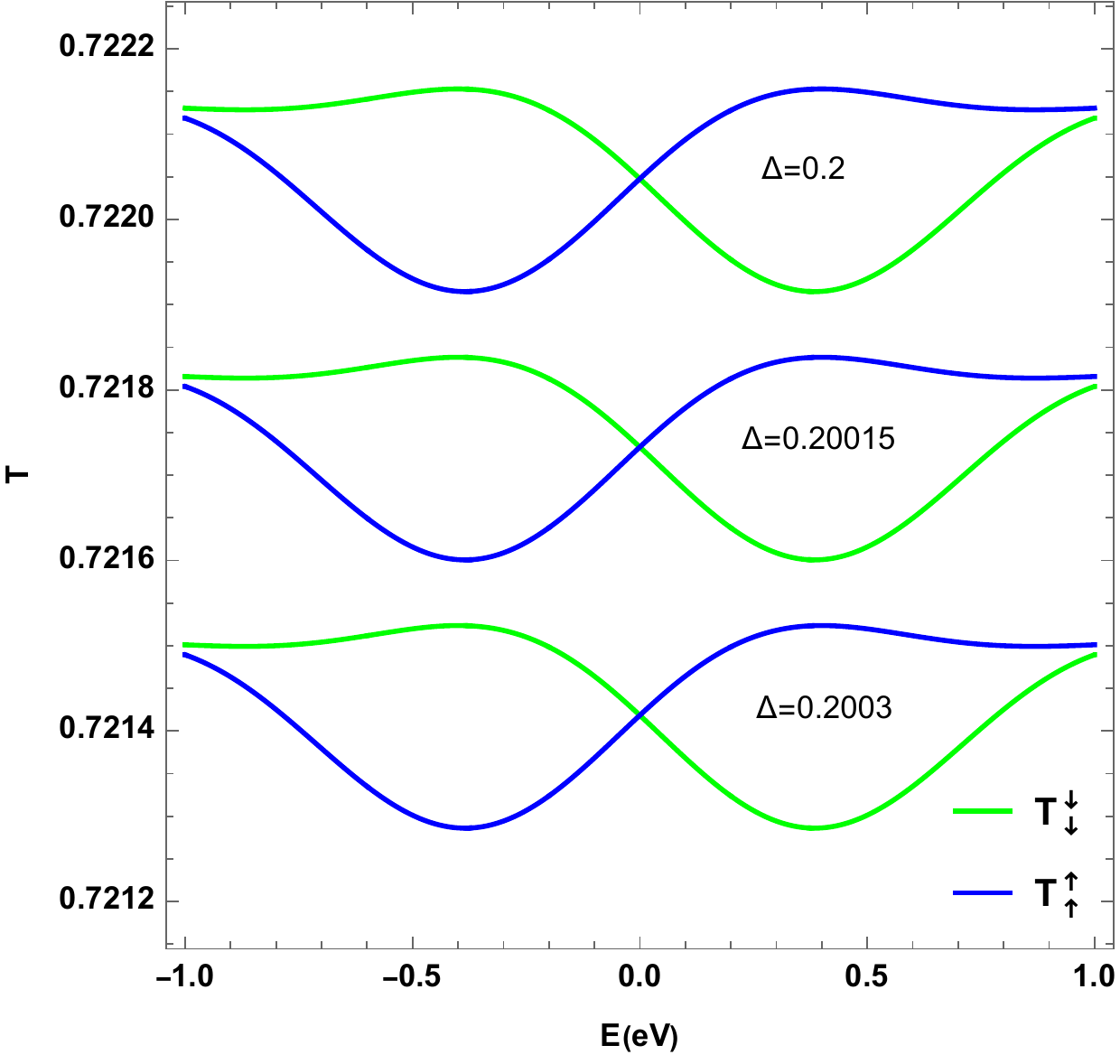}
				\label{subfigurecc}}\\
			\subfloat[]{
				\includegraphics[scale=0.42]{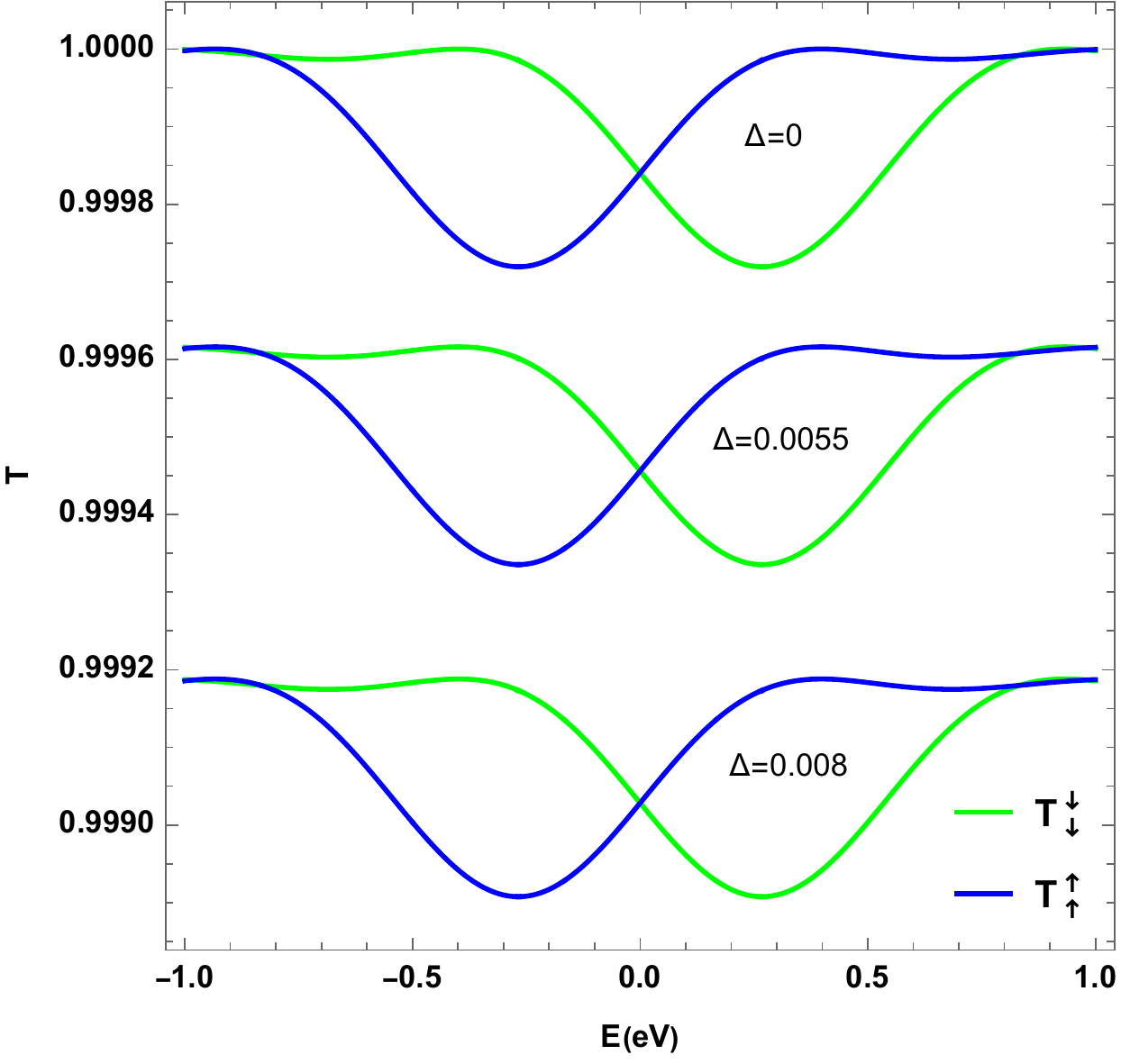}
				\label{subfiguredd}}
			\subfloat[]{
				\includegraphics[scale=0.42]{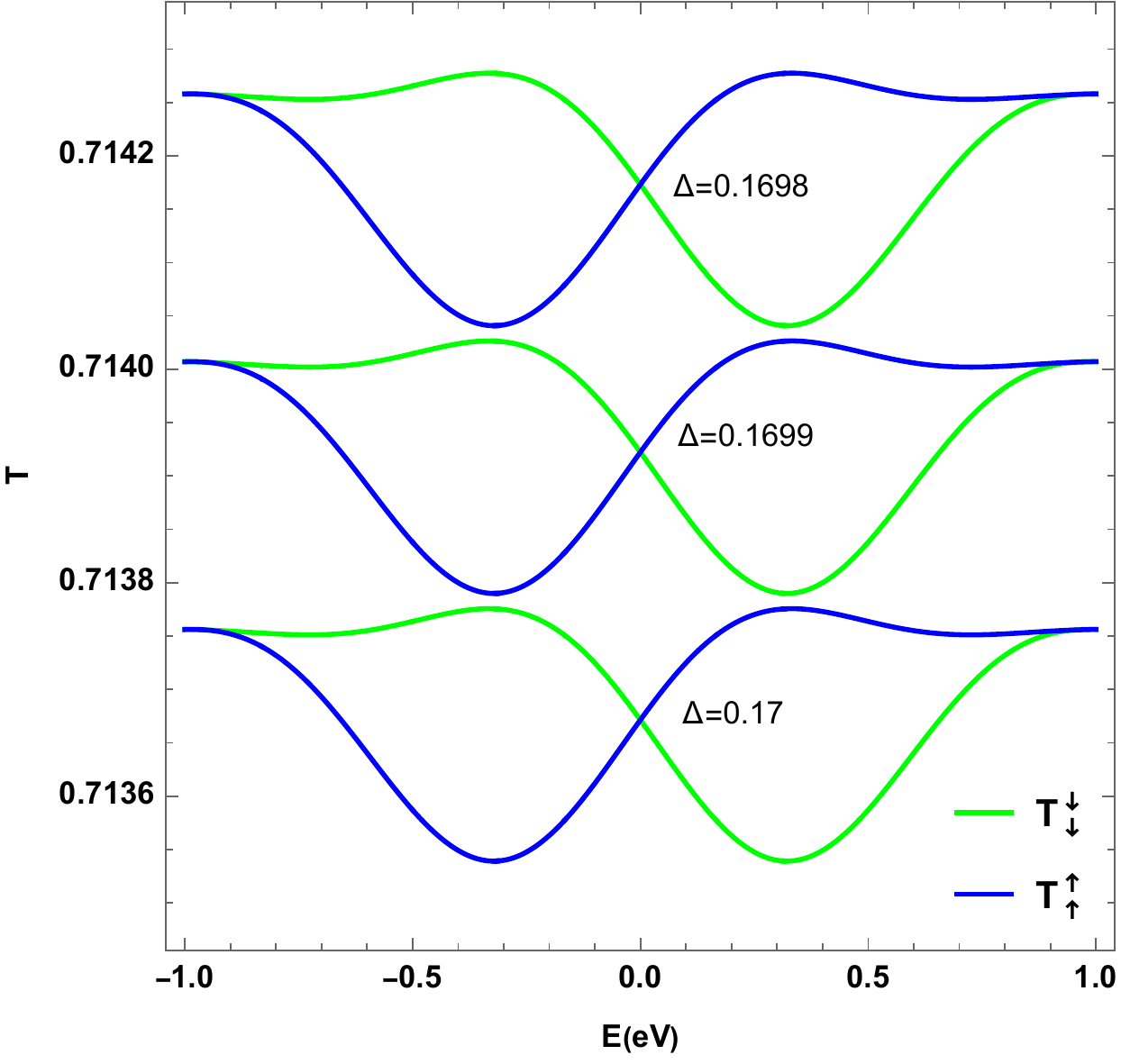}
				\label{subfigureee}}
			\subfloat[]{
				\includegraphics[scale=0.42]{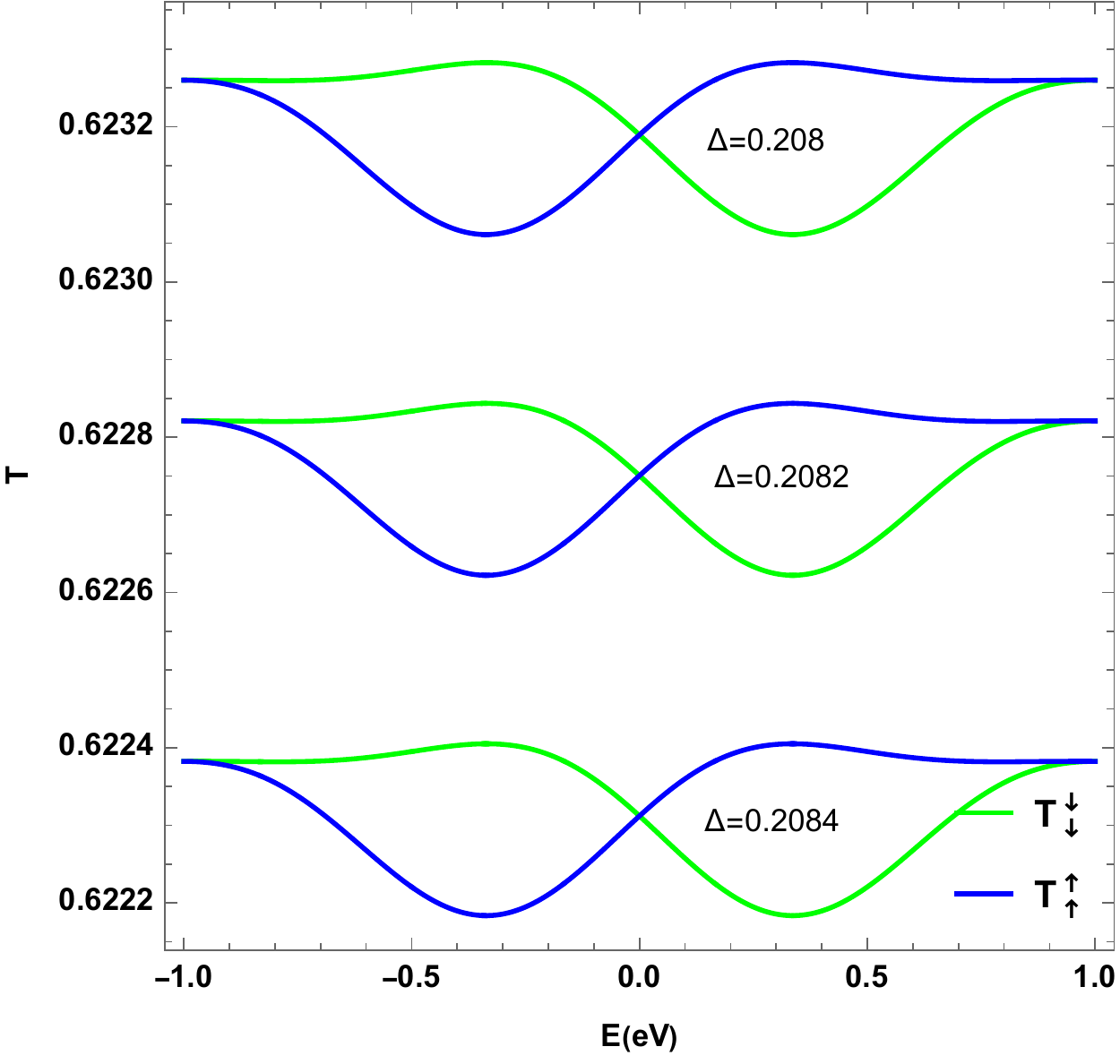}
				\label{subfigureff}}
		\end{center}
		\caption{(color online) The transmission probabilities $T_{\uparrow}^{\uparrow}$ (blue) and $T_{\downarrow}^{\downarrow}$ (green) at normal incidence as a function of the incident energy $E$ for different values of the band gap  $\Delta$ with $\phi=\frac{4\pi}{5}$. 
			The radius $r_{0}=10$ $ \angstrom$ for top panels and  $r_{0}=12 $ $\angstrom$ for bottom panels.}
		\label{delta1}              
	\end{figure}

	We study 
	the tunneling properties of  our system under suitable choices of  the involved   parameters ($r_0, \Delta, \phi $)
	at  normal incidence, i.e. $k_{y}=0$. Indeed, 
	Figure~\ref{delta1} shows 
	the transmissions $ T_{\uparrow}^{\uparrow} $ (blue) and $ T_{\downarrow}^{\downarrow} $ (green) with same spin  as a function of the incident energy $E$ 
	for different values of the band gap with the angle $\phi=\frac{4\pi}{5}$. 
	As a first result we notice that the transmission preserves the symmetry because we have
	$T_{\uparrow}^{\uparrow}(E)=T_{\downarrow}^{\downarrow}(-E)$ for all $E$.
	Now 
	in the upper panels 
	we choose the radius of curvature $r_{0}=10$ $  \angstrom$, then for  $\Delta = 0$ one observes that the transmission  behavior varies slowly closed to the unit.
	As for $\Delta\neq 0$, according to  Figures~\ref{subfigureaa},\ref{subfigurebb},\ref{subfigurecc} we notice that the transmission decreases as long as  $\Delta$ increases and {starts} to move away from the unit. Furthermore, the increase of  $\Delta$ causes a delay in the energy interval where the transmission of the spin down becomes minimal at $E=\sqrt {\lambda_{x}^2 + \Delta^2}$. The bottom panels are as before except that 
	$r_{0} = 12$ $ \angstrom$ and then Figures~\ref{subfiguredd},\ref{subfigureee},\ref{subfigureff} tell us  that the transmission decreases by increasing $\Delta$ while  its symmetrical shape still the same. 
	
\begin{figure}[ht]
		\begin{center}
			\subfloat[]{
				\includegraphics[scale=0.42]{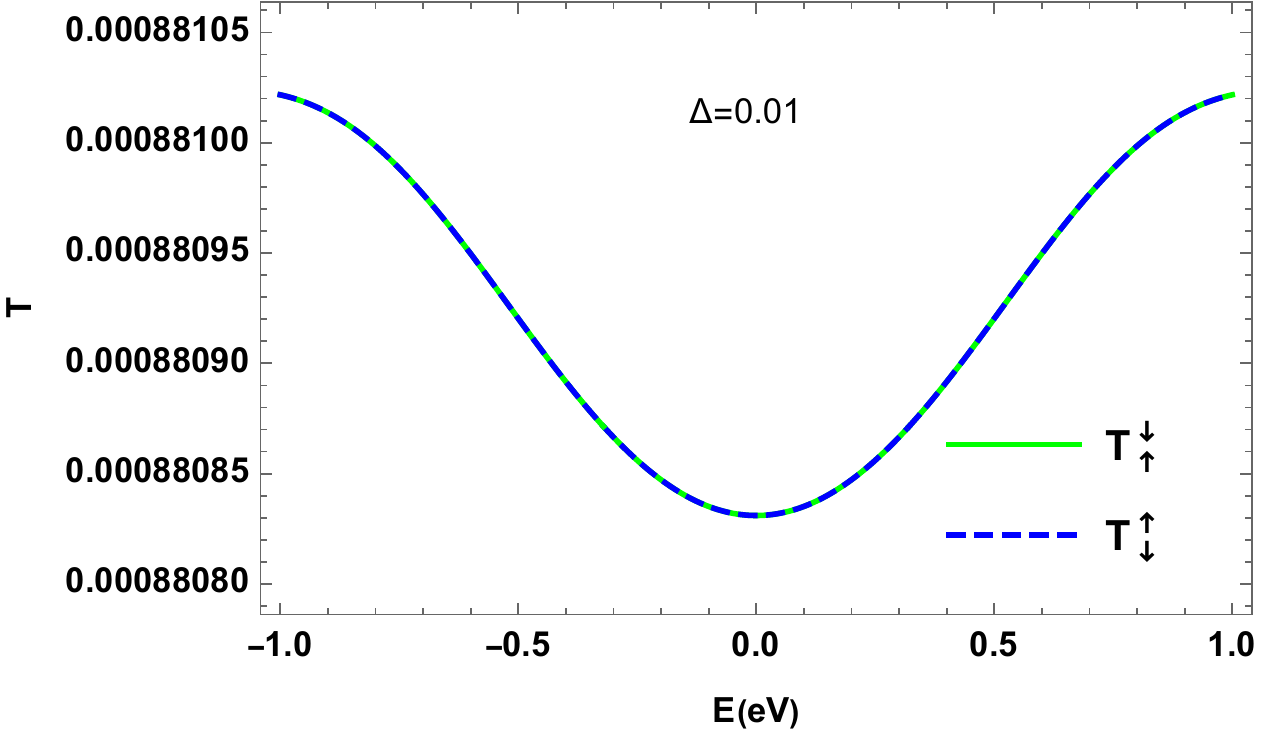}
				\label{subfigureua}}
			\subfloat[]{
				\includegraphics[scale=0.42]{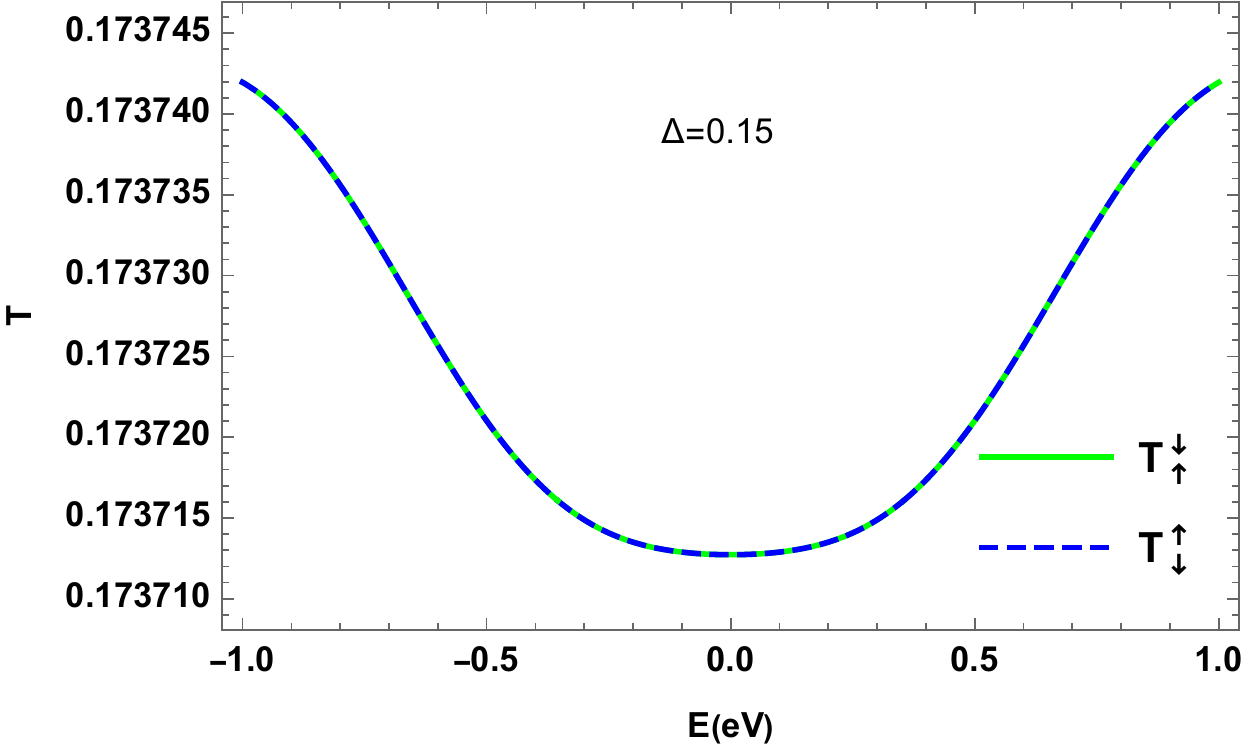}
				\label{subfigureub}}
			\subfloat[]{
				\includegraphics[scale=0.42]{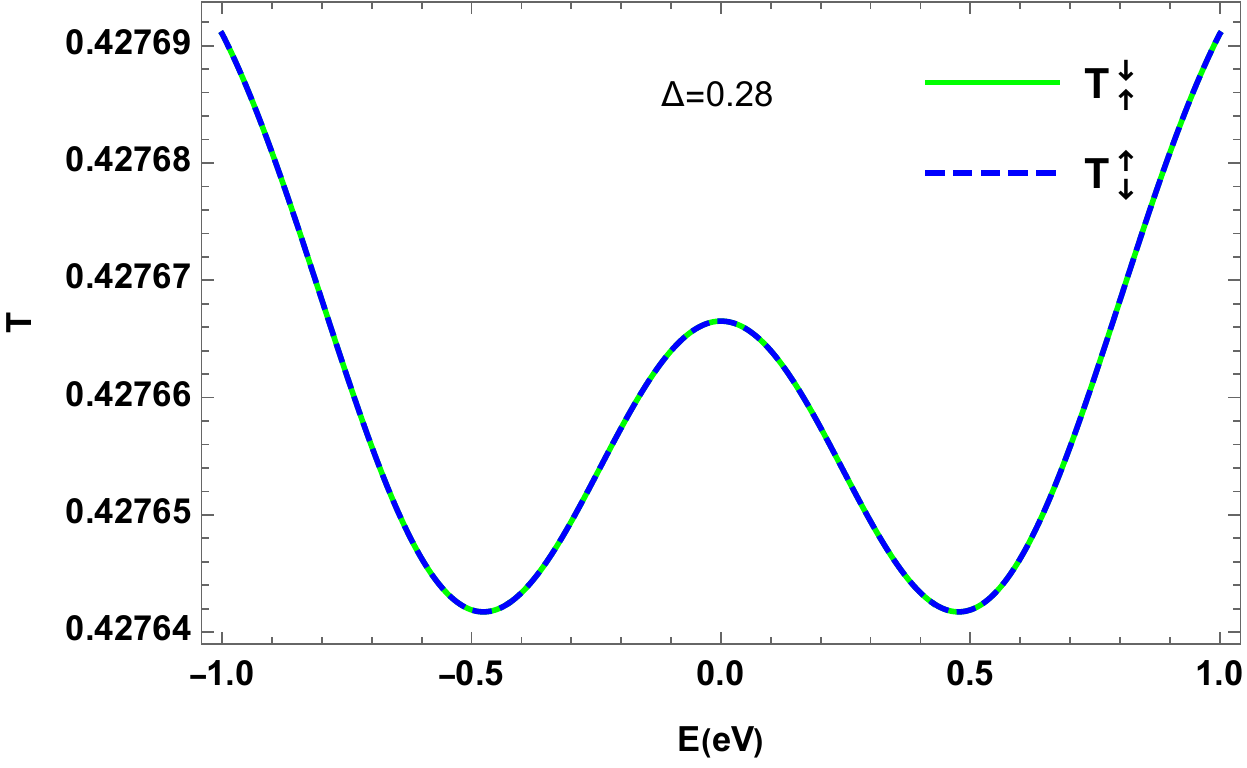}
				\label{subfigureuc}}\\
			\subfloat[]{
				\includegraphics[scale=0.42]{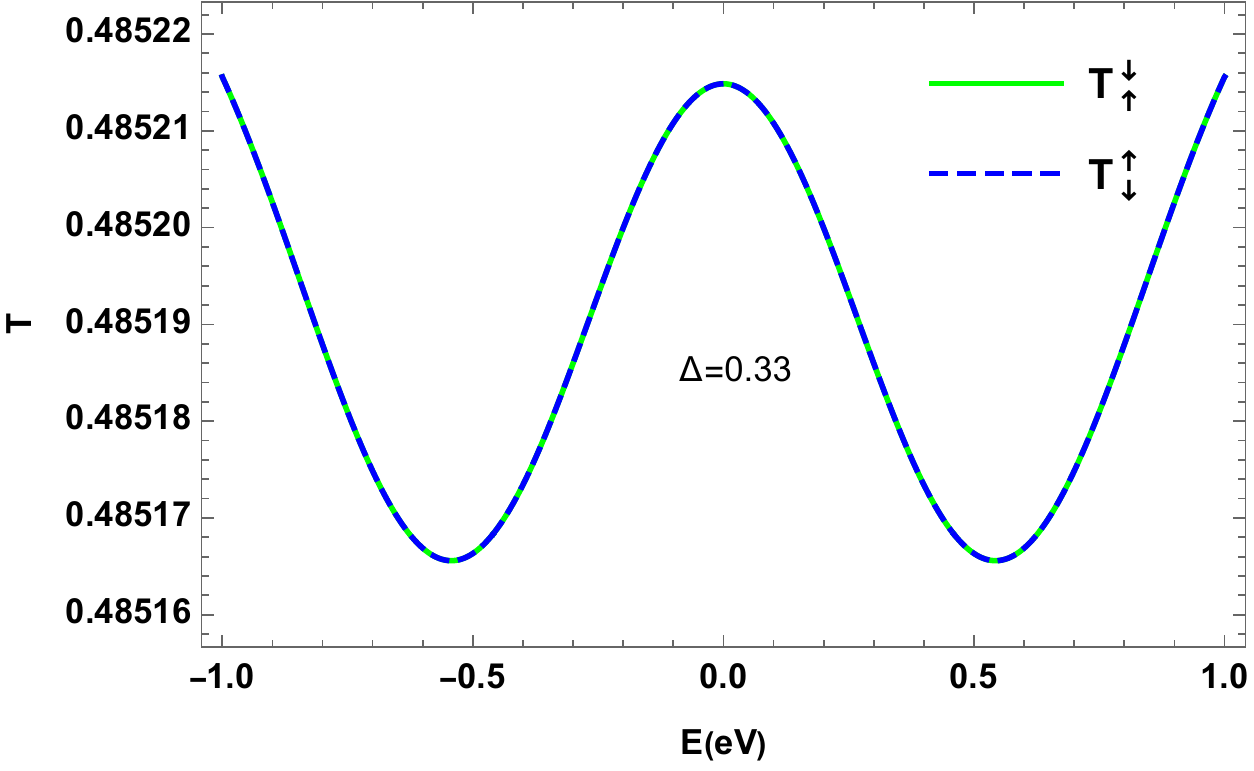}
				\label{subfigureud}}
			\subfloat[]{
				\includegraphics[scale=0.42]{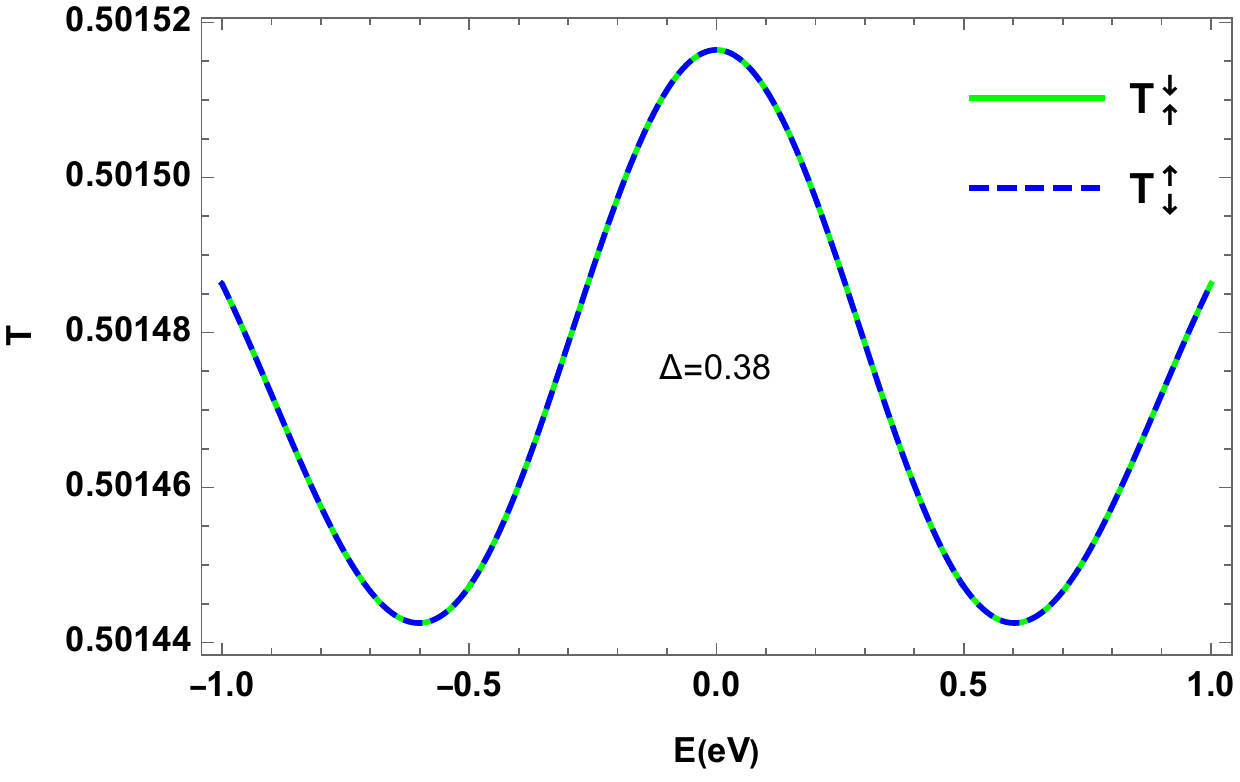}
				\label{subfigureue}}
			\subfloat[]{
				\includegraphics[scale=0.42]{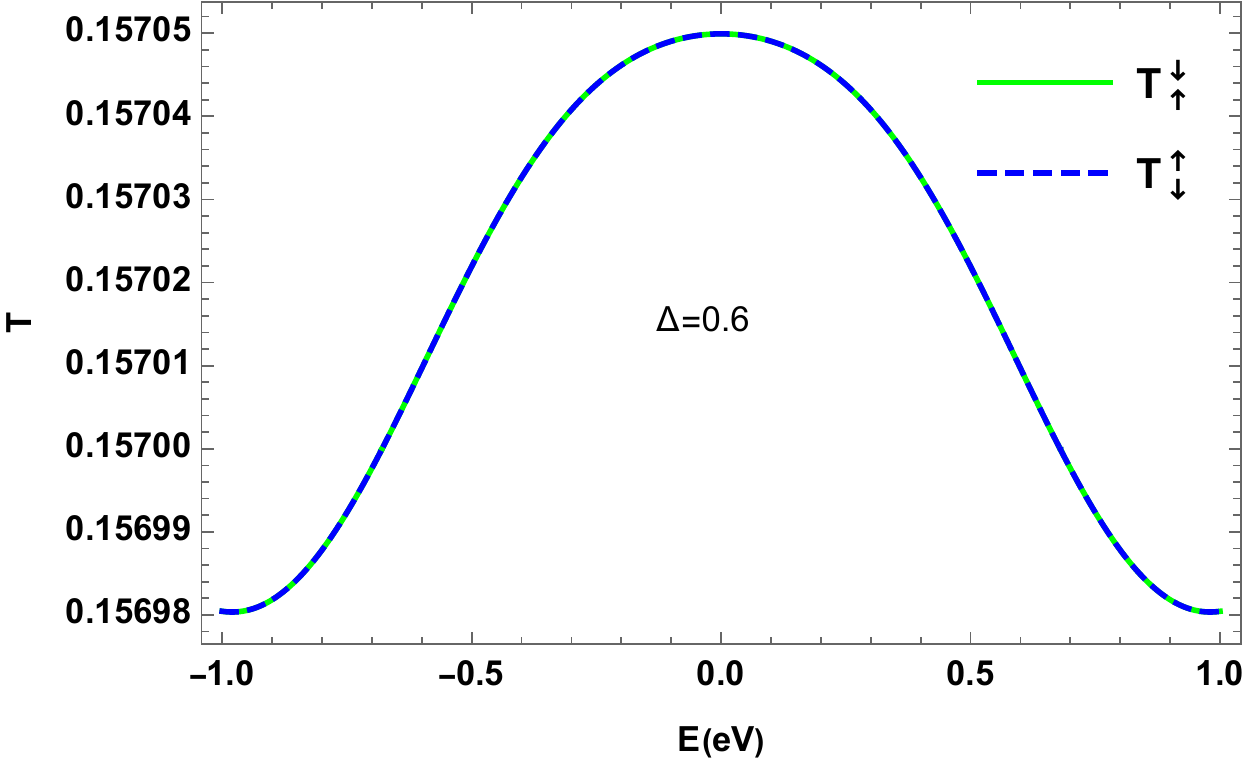}
				\label{subfigureuf}}
		\end{center}
		\caption{(color online) The transmission probabilities $T^{\uparrow}_{\downarrow}$
			(dashed blue) and $ T^{\downarrow}_{\uparrow}$ (green) at normal incidence as a function of the incident  energy $E$ for different values of band gap $\Delta$
			with $\phi=\frac{4\pi}{5}$ and $r_{0}=10$ $\angstrom.$}
		\label{delta2}              
	\end{figure}

	
	In Figure~\ref{delta2} we present the transmissions of the opposite spin 
	$T^{\uparrow}_{\downarrow}$ (dashed blue) and $ T^{\downarrow}_{\uparrow}$ (green)
	as a function of the incident energy $E$ for  different values of the band gap $\Delta$ with the angle $\phi=\frac{4\pi}{5}$ and radius $r_{0}=10\ \angstrom$. 
	We emphasis that both transmissions always keep a symmetrical behavior such as the relation $T_{\uparrow}^{\downarrow}= T_{\downarrow}^{\uparrow}$ is satisfied. We start by noting that 
	for  $\Delta=0$ in Figure~\ref{subfigureua}  both transmissions are almost null as obtained in \cite{pudlak2015cooperative}. However, for $\Delta\neq0$ it is clear that  the transmission increases  progressively
	by oscillating  under the increase of  $\Delta$ as in particular for  $\Delta=0.15$ eV  in Figure \ref{subfigureub} and $\Delta=0.28$ eV in \ref{subfigureuc}. Interestingly, 
	for $\Delta=0.33$ eV one sees that the transmission can be approached by a
	sinusoidal function as depicted in Figure~\ref{subfigureud} but for $ \Delta=0.38 $ eV
	it oscillates differently according to Figure~\ref{subfigureue} with a remarkably  change in its amplitudes and such manifestation is due of  course to the presence of  $ \Delta$. Finally for $ \Delta=0.6 $ eV
	in Figure~\ref{subfigureuf}, we notice that  the transmission takes a characteristic form that has a Gaussian shape. 
	As a result,
	we observe that the increase of $\Delta$ changes dramatically the transmission behavior.
	It is clearly seen that  $\Delta$ affects all transmission channels and then can be used a controllable toy to adjust the tunneling properties toward a technological application.
	\begin{figure}[H]
		\begin{center}
			\subfloat[]{
				\includegraphics[scale=0.47]{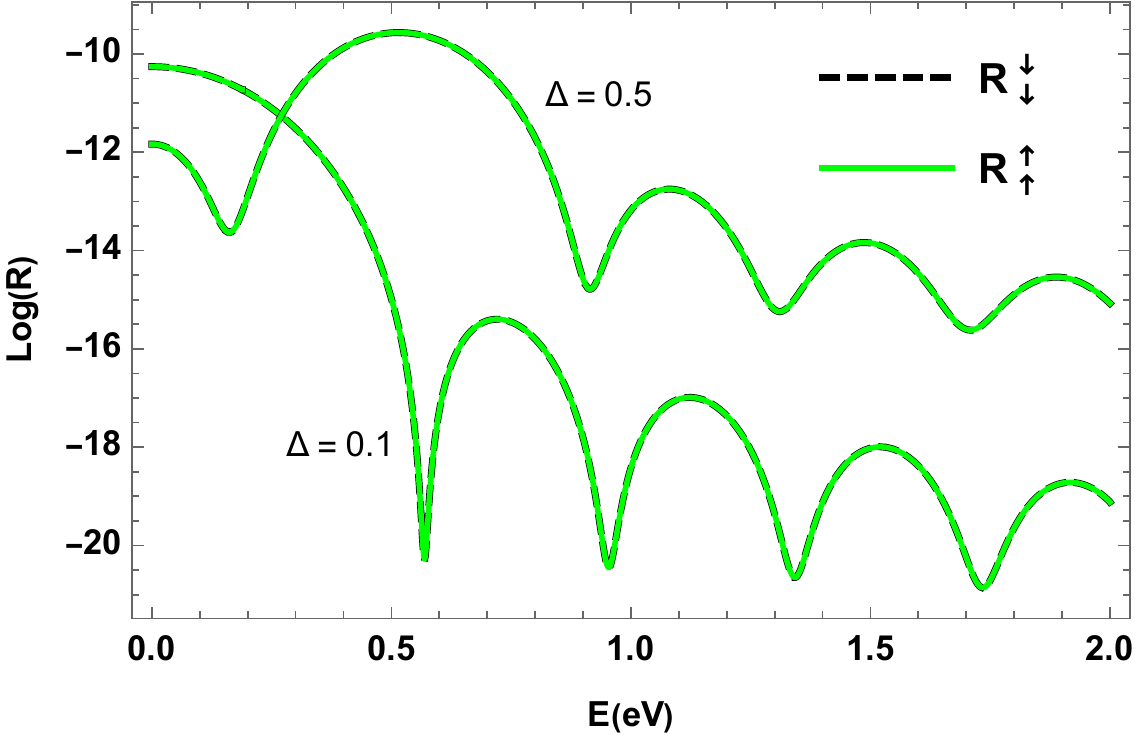}
				\label{subfigurea}}
			\subfloat[]{
				\includegraphics[scale=0.47]{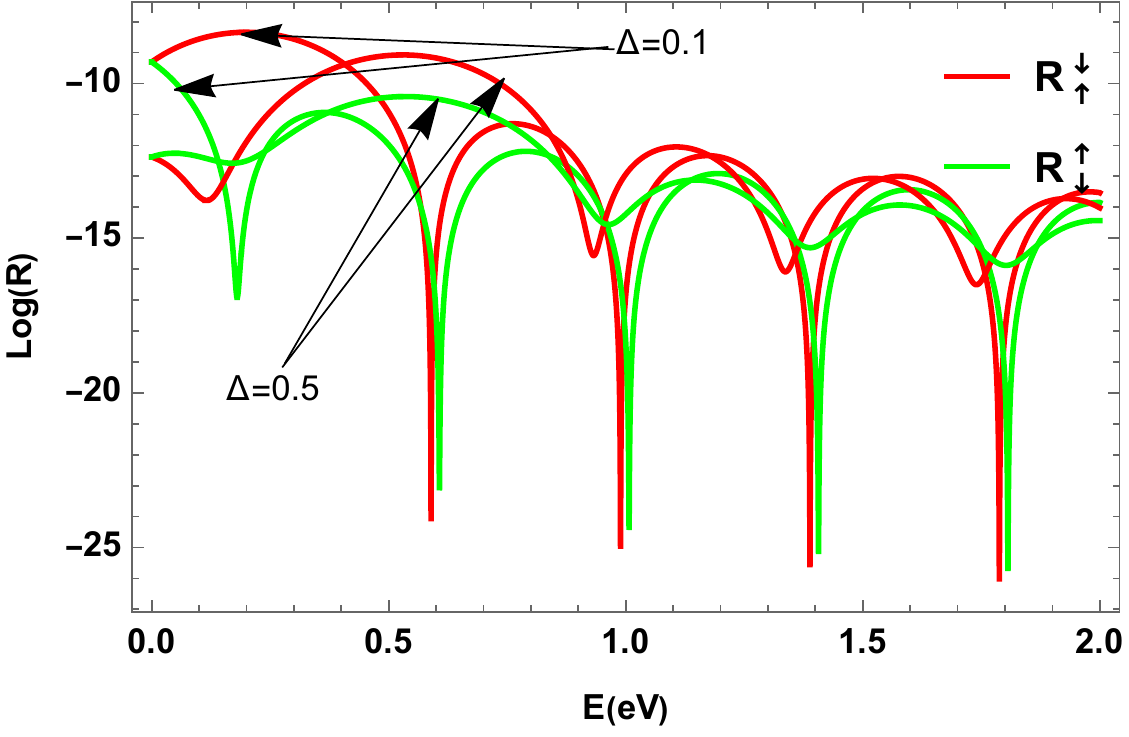}
				\label{subfigureb}}
			\subfloat[]{
				\includegraphics[scale=0.47]{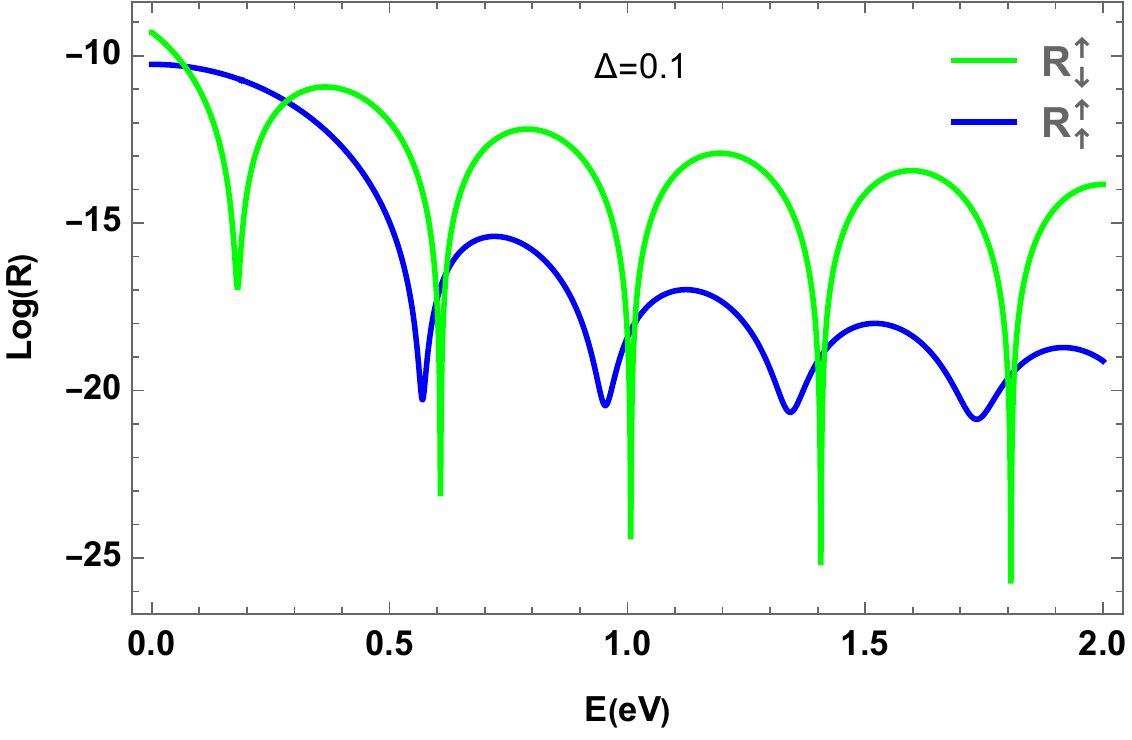}
				\label{subfigurec}}\\
			\subfloat[]{
				\includegraphics[scale=0.47]{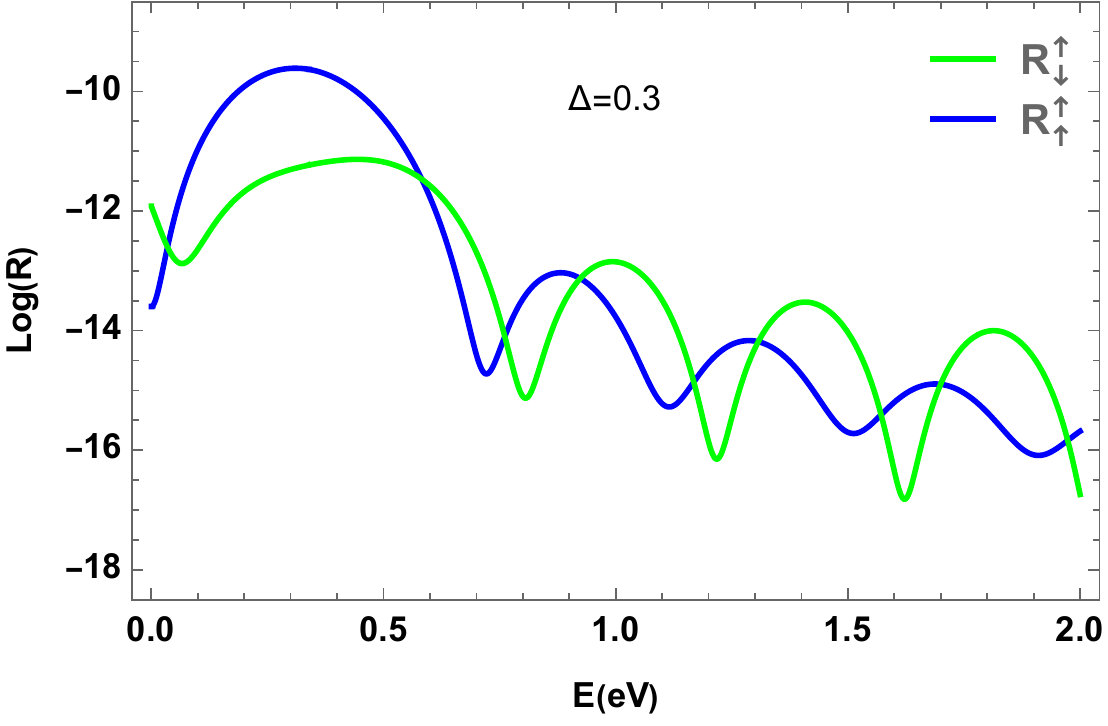}
				\label{subfigured}}
			\subfloat[]{
				\includegraphics[scale=0.47]{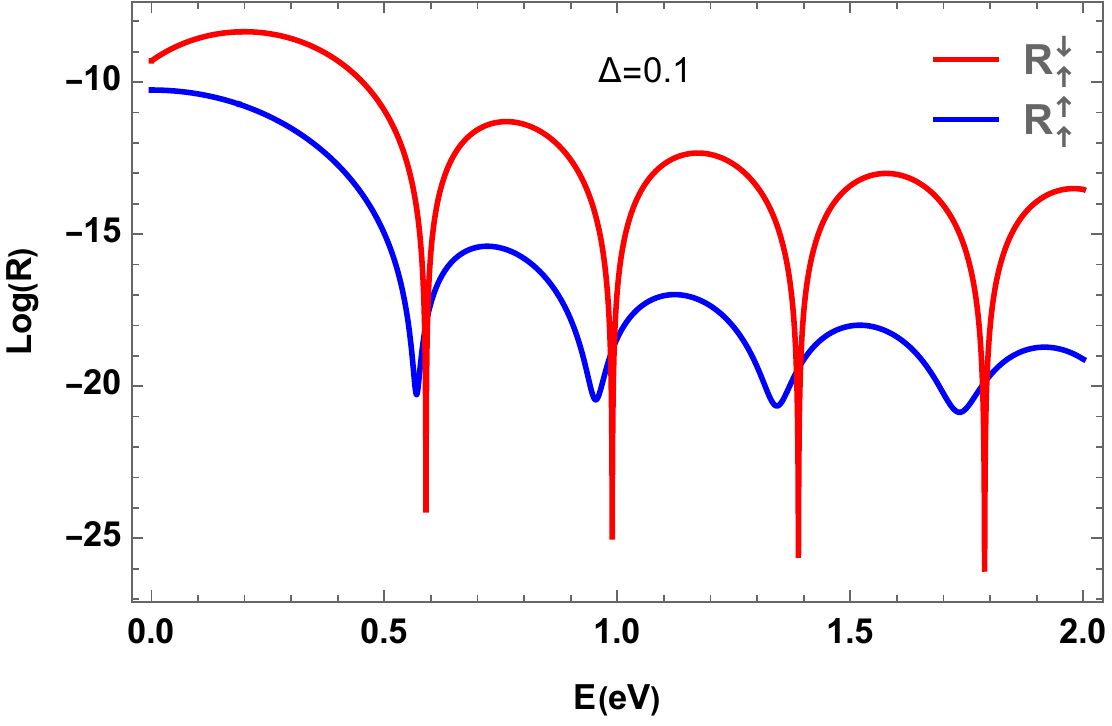}
				\label{subfiguree}}
			\subfloat[]{
				\includegraphics[scale=0.47]{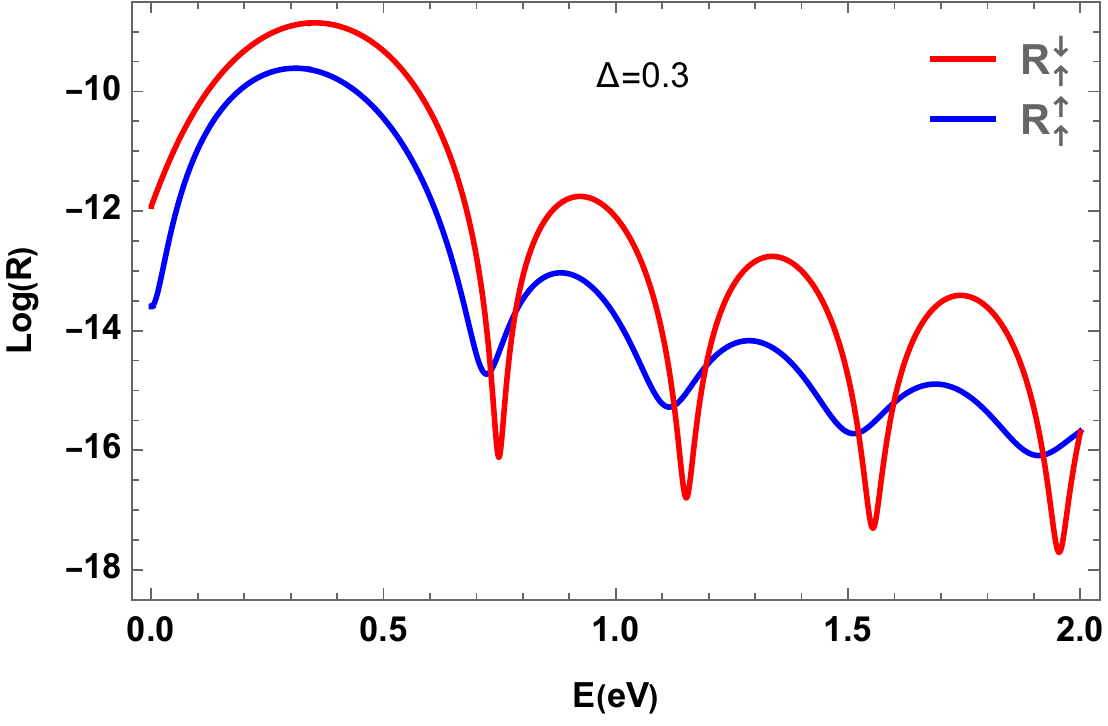}
				\label{subfiguref}}
		\end{center}
		\caption{(color online) The reflection probabilities in the logarithmic scale ($R_{\uparrow}^{\uparrow}=R_{\downarrow}^{\downarrow}$, $R_{\downarrow}^{\uparrow}$, $R_{\uparrow}^{\downarrow}$) at normal incidence as a function of the incident energy $E$ 
			for three values of the band gap $\Delta$  
			with $\phi=\frac{4\pi}{5}$ and  $r_{0}=20\ \angstrom$.}
		\label{delta3}              
	\end{figure}

	In  Figure~\ref{delta3}, we show the reflection probabilities in the logarithmic scale 
	at normal incidence   as a function of the incident  energy $E$ 
	for three values of the band gap $\Delta=(0.1, 0.3, 0.5)$ eV with the ripple angle $\phi=\frac{4\pi}{5}$ and radius $r_{0}=20$ $ \angstrom$.
	As an interesting consequence resulted  from  $\Delta\neq0$ is the emergence of reflections with the same spin, namely
	$R_{\uparrow}^{\uparrow}=R_{\downarrow}^{\downarrow}\neq0$, contrary to what obtained
	in \cite{pudlak2015cooperative} for gapless case  ($\Delta=0$), see  Figure~\ref{subfigurea}. This means that the Klein tunneling is not always satisfied  with the presence of band gap. Figure~\ref{subfigureb} tells us that the reflections with opposite spin do not show the same behavior and consequently we have $R_{\uparrow}^{\downarrow}\neq R_{\downarrow}^{\uparrow}$. 
	By comparing 	$R_{\uparrow}^{\uparrow}$ and $R_{\downarrow}^{\uparrow}$ 
	we observe that 
	the sharp peaks appearing in Figure~\ref{subfigurec} for $\Delta=0.1$ eV
	disappear in Figure \ref{subfigured} for $\Delta=0.3$ eV
	and then oscillations modes take place. We have the same conclusion 
	for 	$R_{\uparrow}^{\uparrow}$ and $R^{\downarrow}_{\uparrow}$
	in  Figure~\ref{subfiguree} for $\Delta=0.1$ eV
	disappear in Figure \ref{subfiguref} for $\Delta=0.3$ eV. Indeed,  
	we
	observe that there are resonances in reflection with different amplitudes, which become important for  $R_{\downarrow}^{\uparrow}$ and $R_{\uparrow}^{\downarrow}$ in particular  for $\Delta=0.1$ eV as presented in Figure~\ref{subfiguree}. This behavior changes for 
	$\Delta=0.3$ in Figure~\ref{subfiguref} where all reflection channels oscillate with large amplitudes and less resonances. 

	\begin{figure}[H]
		\begin{center}
			\subfloat[]{
				\includegraphics[scale=0.4]{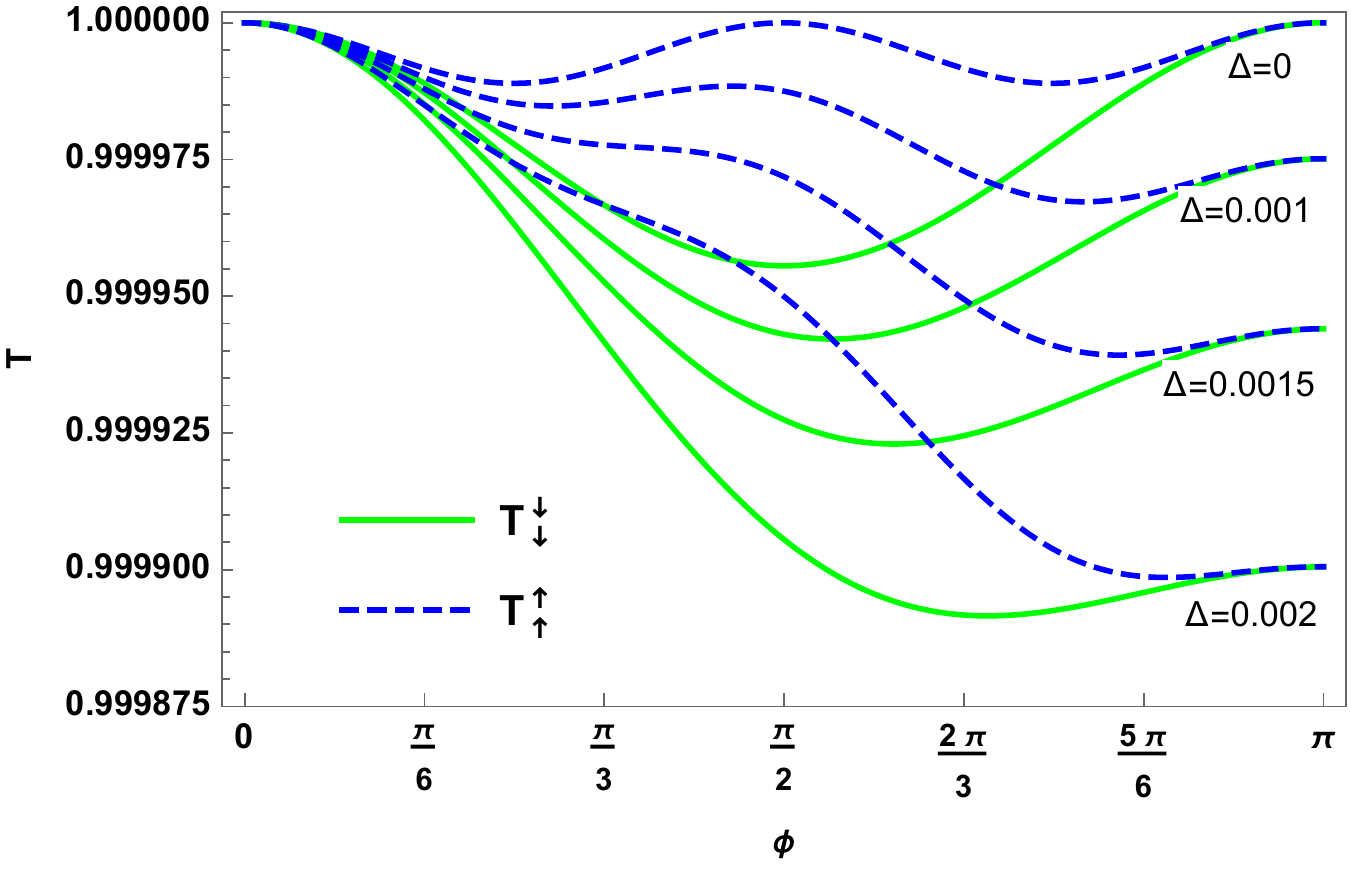}
				\label{subfiga}}
			\subfloat[]{
				\includegraphics[scale=0.4]{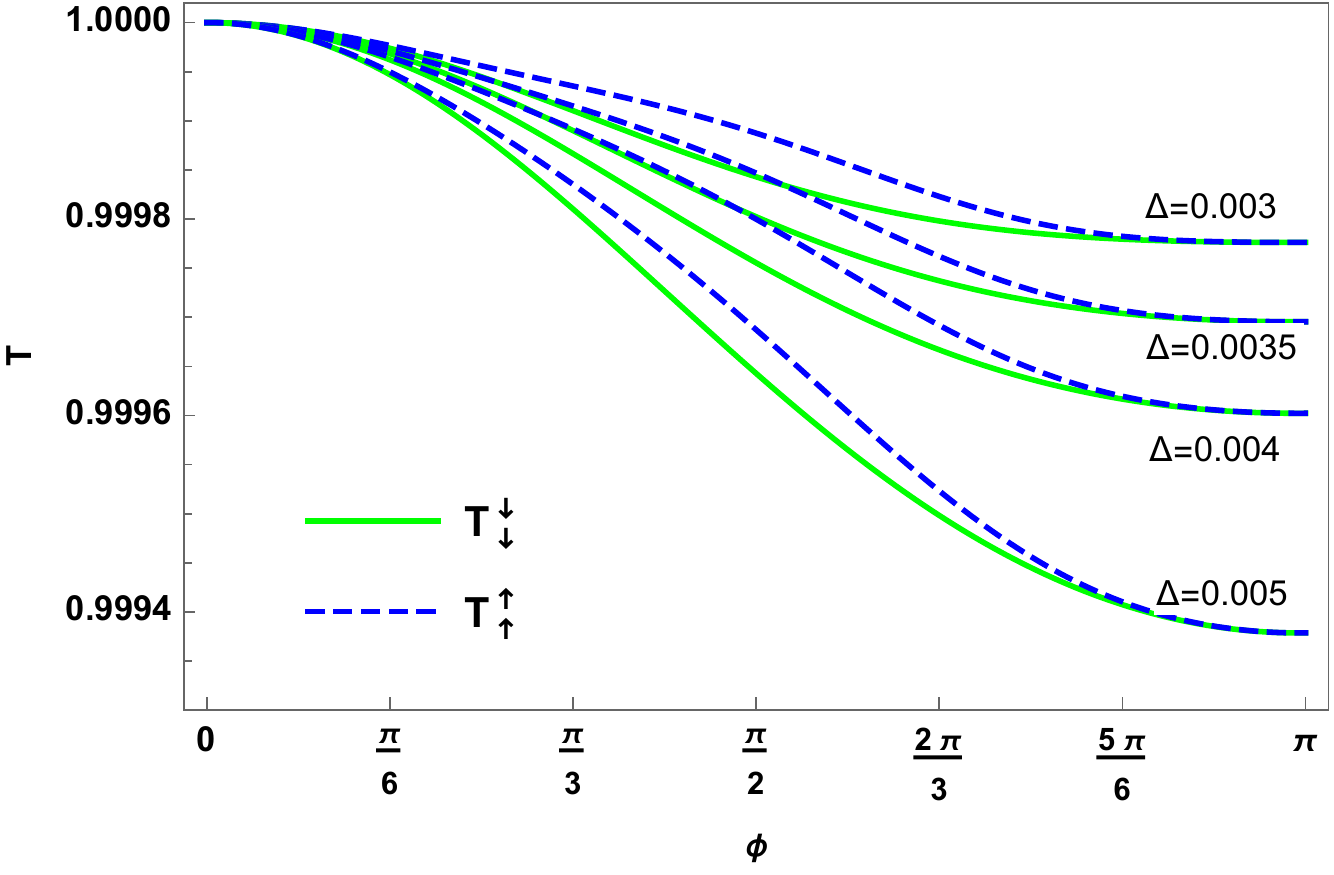}
				\label{subfigb}}
			\subfloat[]{
				\includegraphics[scale=0.4]{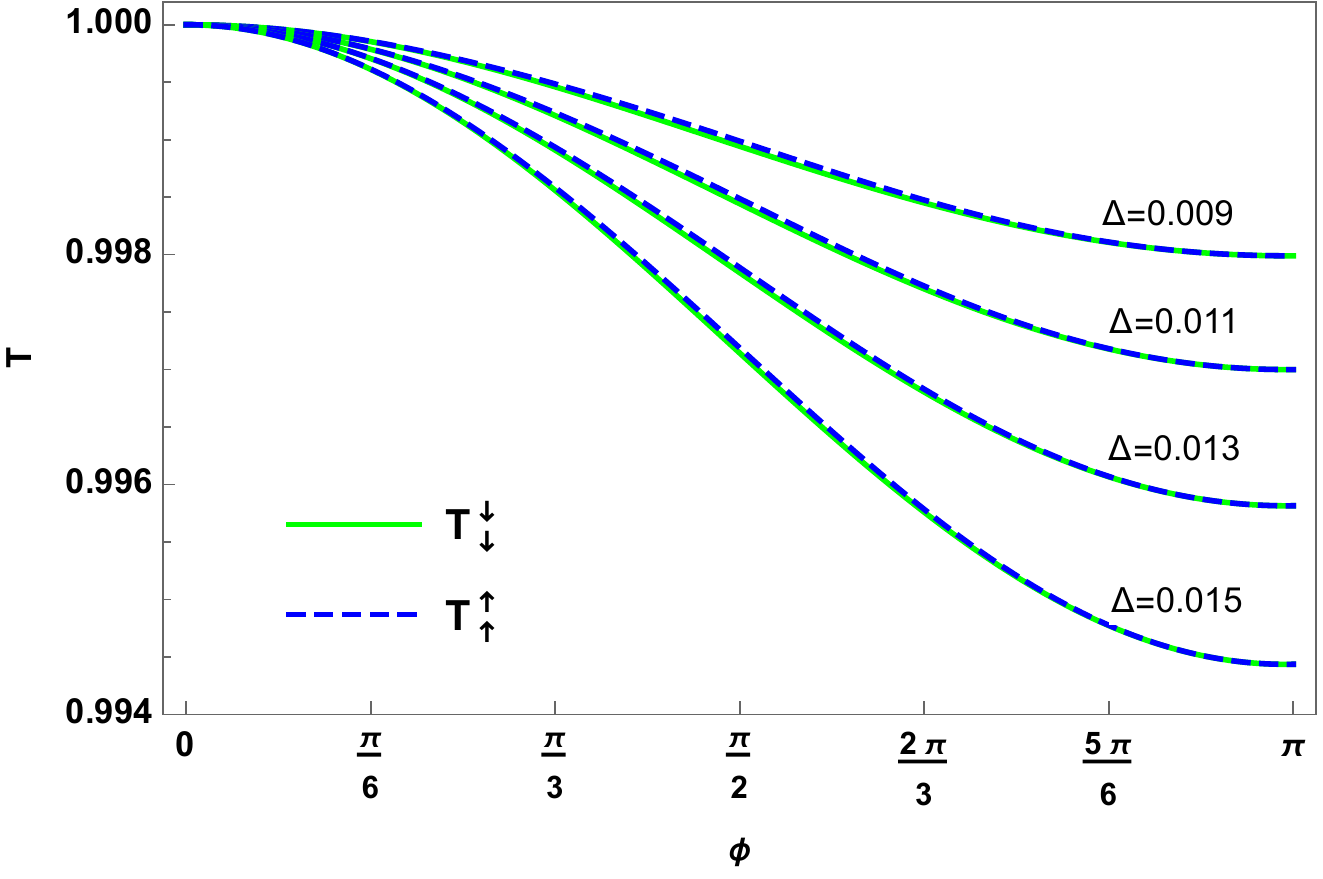}
				\label{subfigc}}\\
			\subfloat[]{
				\includegraphics[scale=0.4]{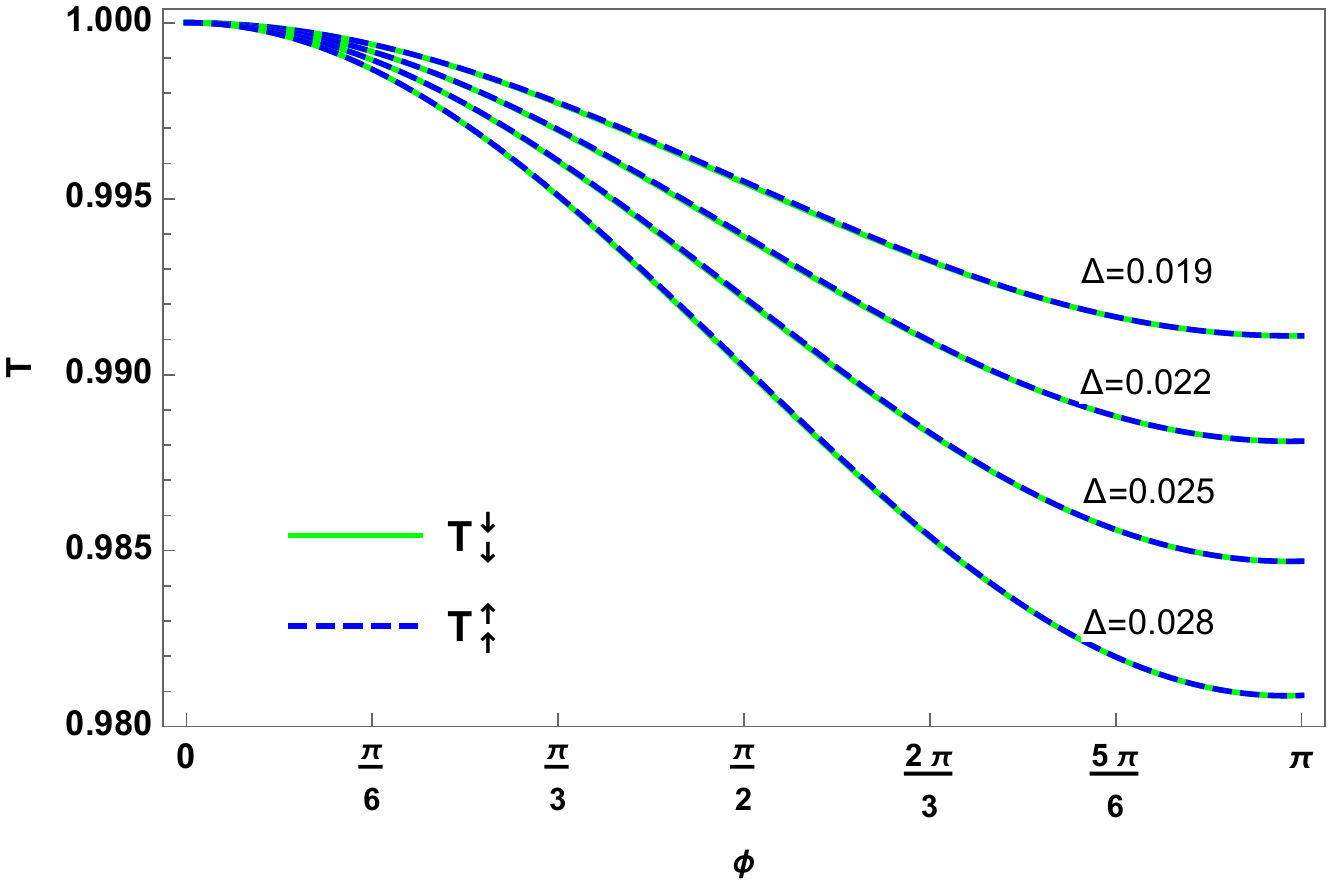}
				\label{subfigd}}
			\subfloat[]{
				\includegraphics[scale=0.4]{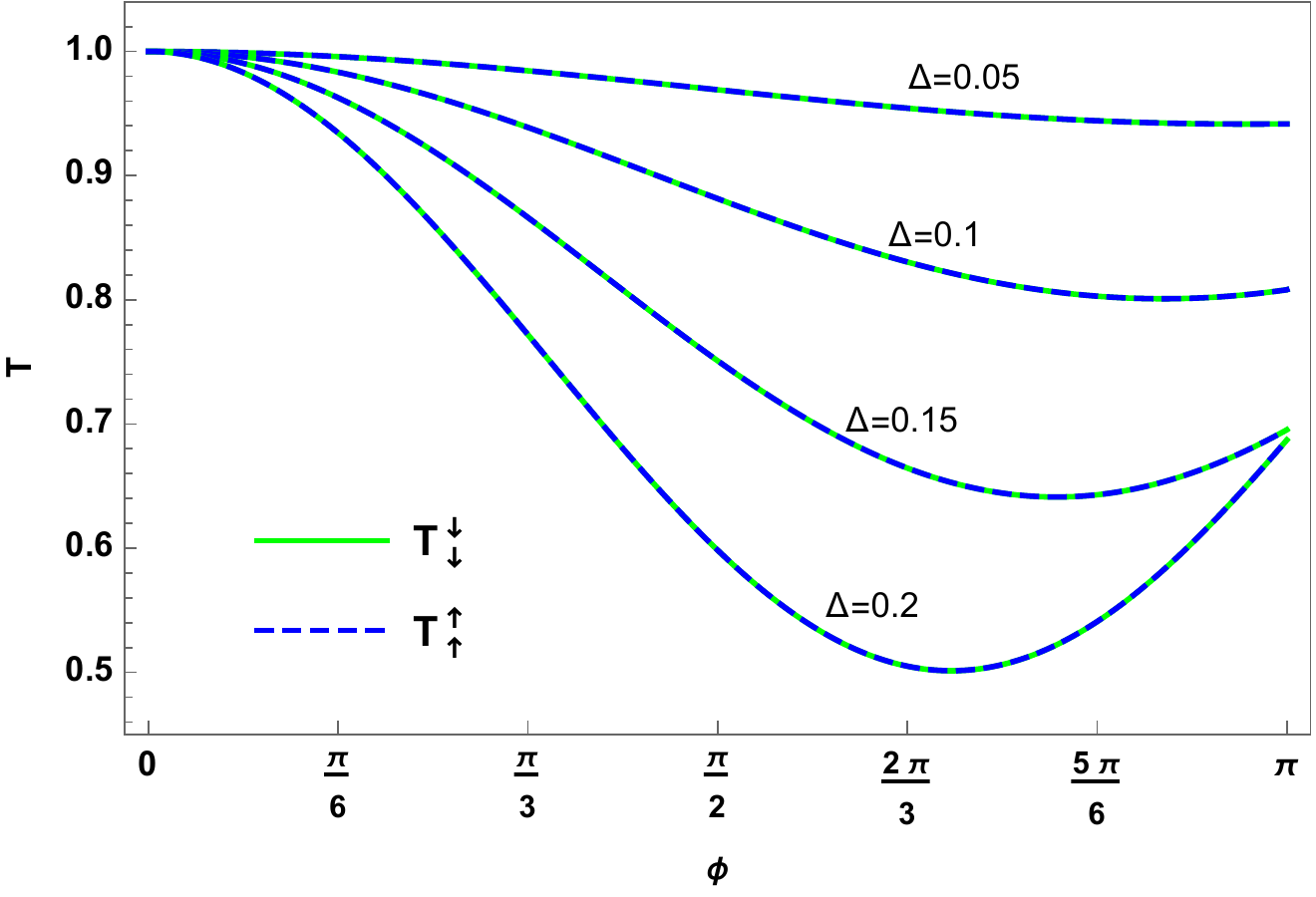}
				\label{subfige}}
			\subfloat[]{
				\includegraphics[scale=0.4]{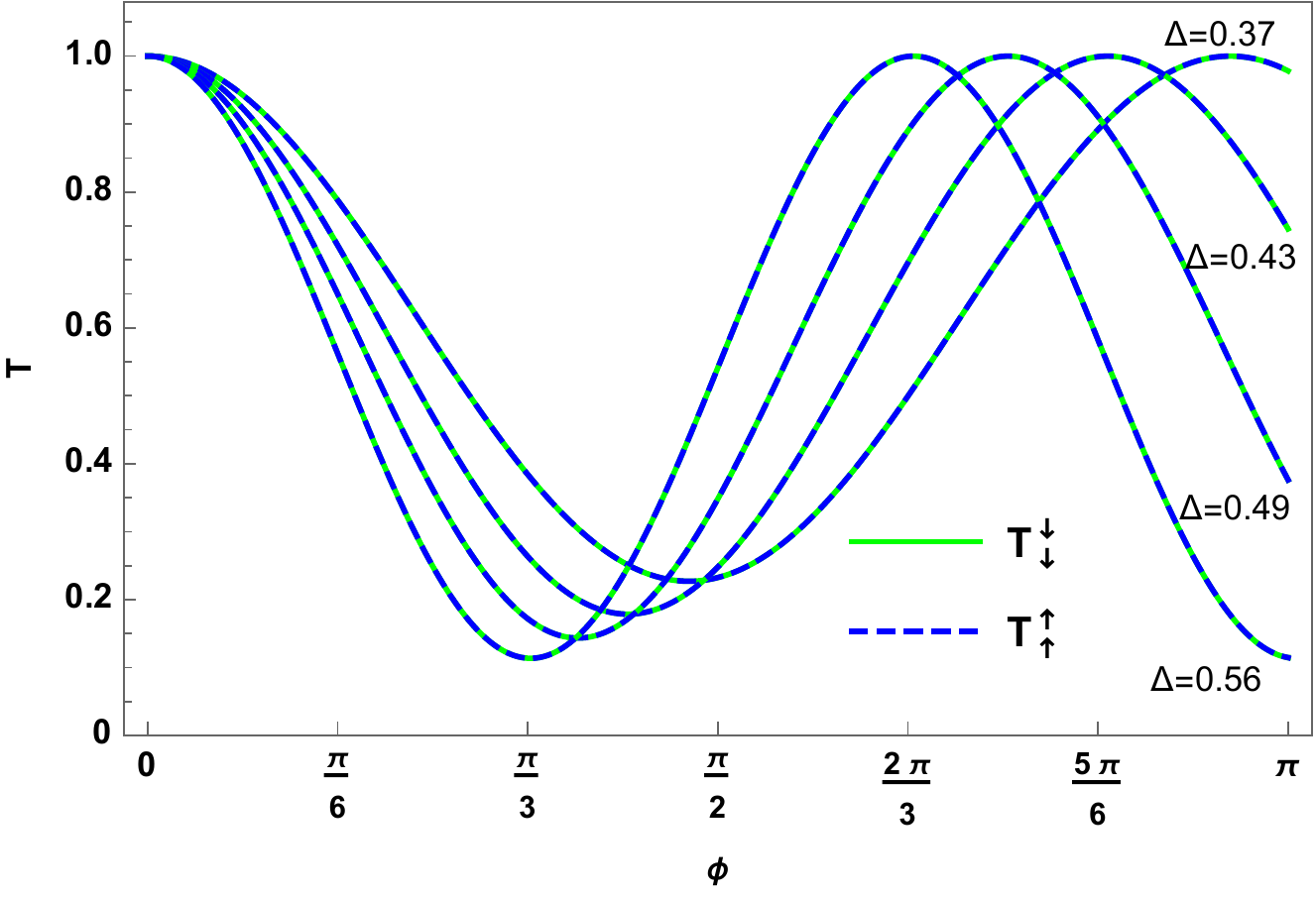}
				\label{subfigf}}
		\end{center}
		\caption{(color online) The transmission probabilities $T_{\uparrow}^{\uparrow}$ (blue dashed lines) and $T_{\downarrow}^{\downarrow}$ (green solid lines) at normal incidence as a function of the angle $\phi$ of ripple  for 
			different values of the band gap $\Delta$  
			with $E=0.6$ 
			eV
			and $r_{0}=16$ $ \angstrom$.}
		\label{delta4}              
	\end{figure}
	
	In  Figure~\ref{delta4}, we show the transmission probabilities with the same spin as a function of the ripple angle $\phi$ for different values of the band gap $\Delta$ with
	incident energy $E = 0.6$ eV  
	and radius $r_{0}=16\ \angstrom$.
	Our results show a degradation of the capacity of spin filtering in our system due to
	$\Delta$. 
	Consequently, in Figure~\ref{subfiga} we observe that $T_{\uparrow}^{\uparrow}$ and $T_{\downarrow}^{\downarrow}$ have  maximum values  at $\phi =\pi$ for the case  $\Delta = 0$ as found in \cite{pudlak2015cooperative}. Now with the inclusion of  mass term, we notice that  
	both of transmissions decrease as long as $\phi$ increases but they approach to each others see Figures~\ref{subfigb},\ref{subfigc}. By increasing $\Delta$, the transmissions coincide and show different behaviors in 
	Figures~\ref{subfigd},\ref{subfige}. In particular, they show periodically oscillations as in Figure~\ref{subfigf} and then one can theoretically approach 
	them by  sinusoidal functions. Consequently, we notice from a critical  value of $\Delta$ the spin splitting is not longer maintained as clearly seen starting from Figure~\ref{subfigc}, which reduces the spin degree of freedom by giving rise only to one transmission channel instead of two.
	Then, one can use $\Delta $ as a key ingredient to control 
	{the spin splitting
	 in our system.}

	
	\begin{figure}[H]
		\begin{center}
			\subfloat[]{
				\includegraphics[scale=0.4]{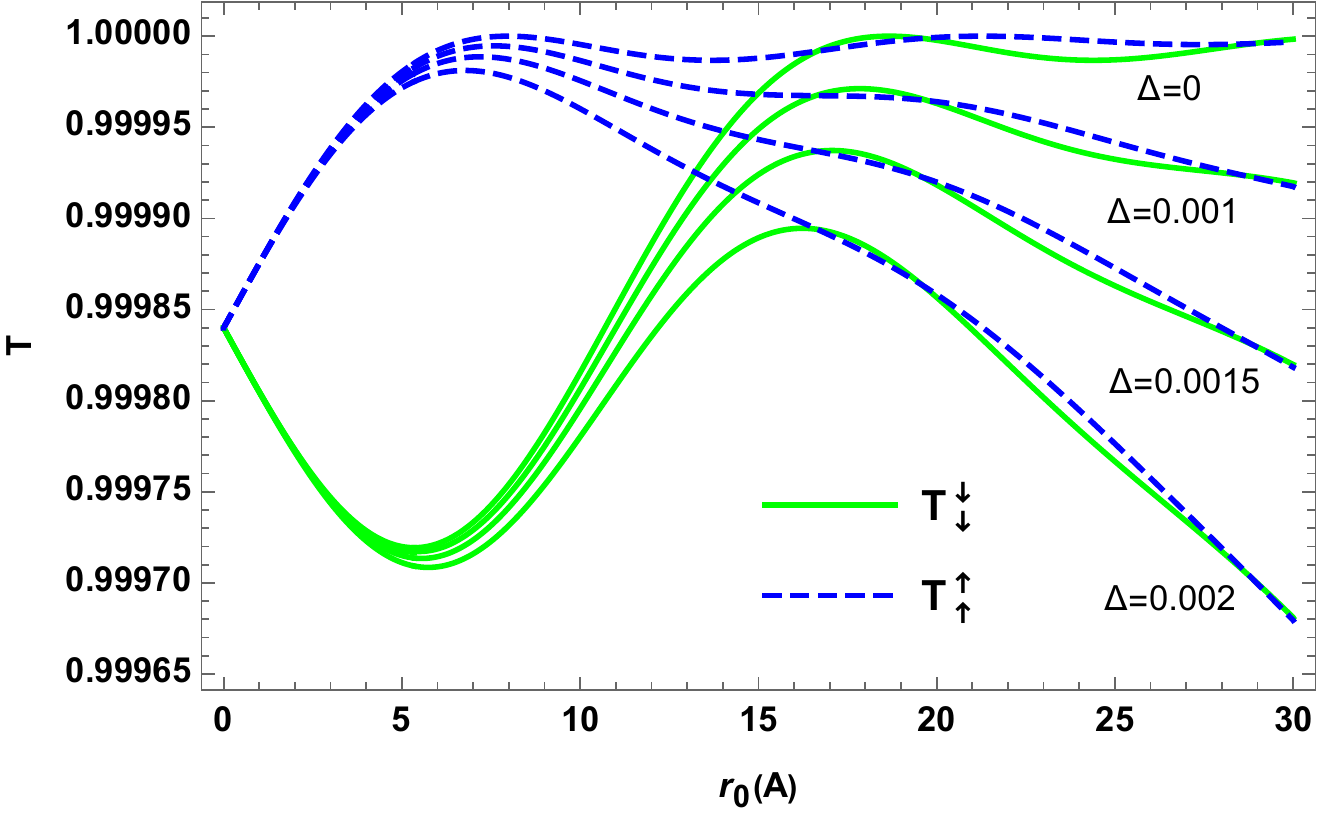}
				\label{subfigatr}}
			\subfloat[]{
				\includegraphics[scale=0.4]{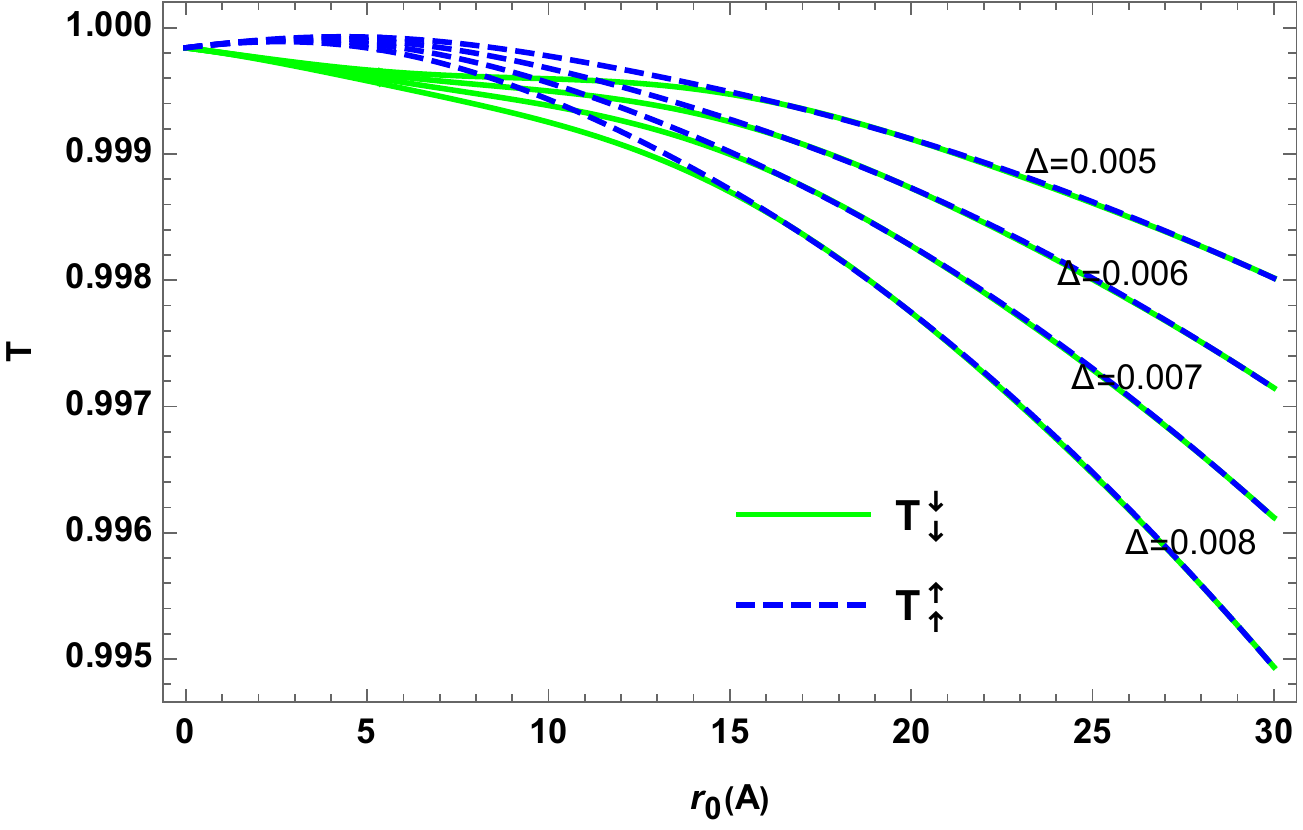}
				\label{subfigbtr}}
			\subfloat[]{
				\includegraphics[scale=0.4]{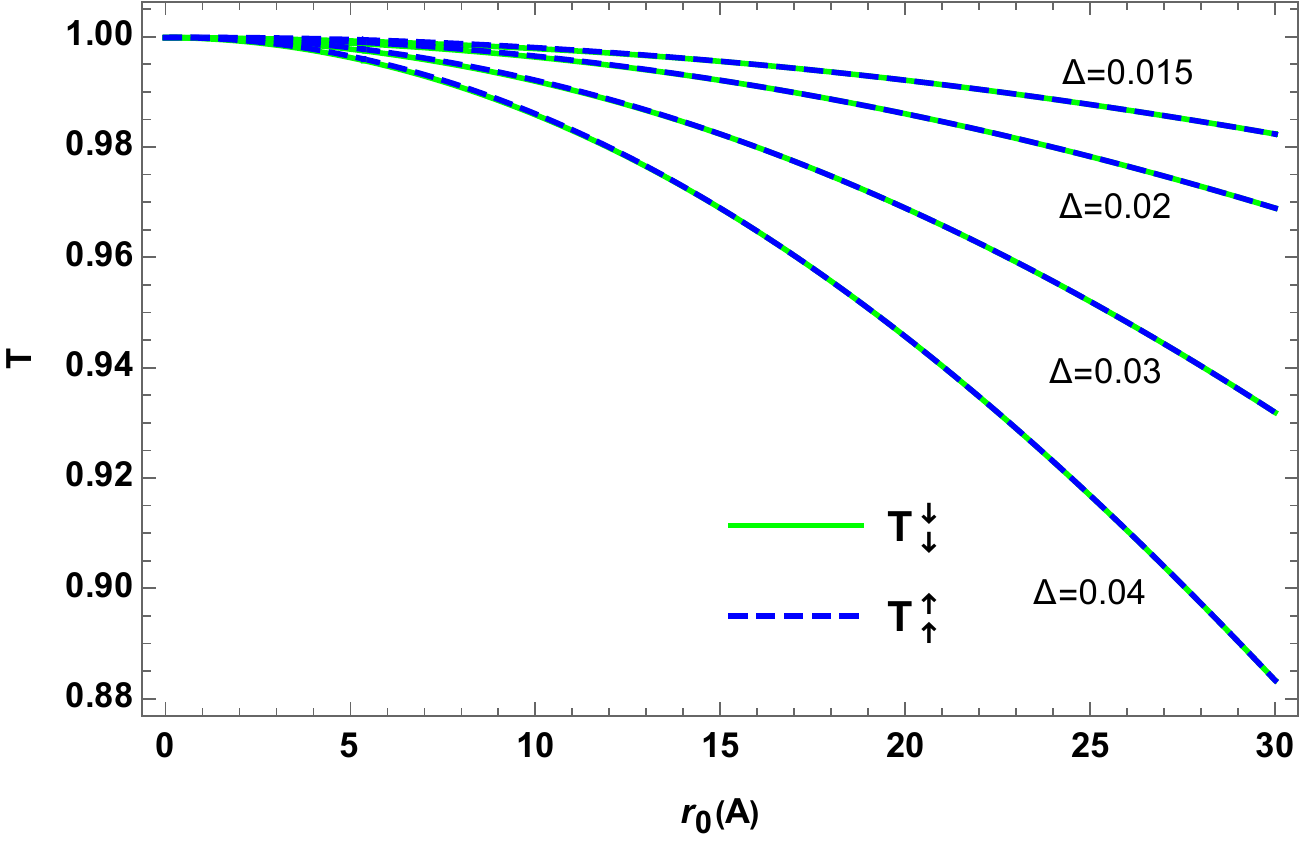}
				\label{subfigctr}}\\
			\subfloat[]{
				\includegraphics[scale=0.4]{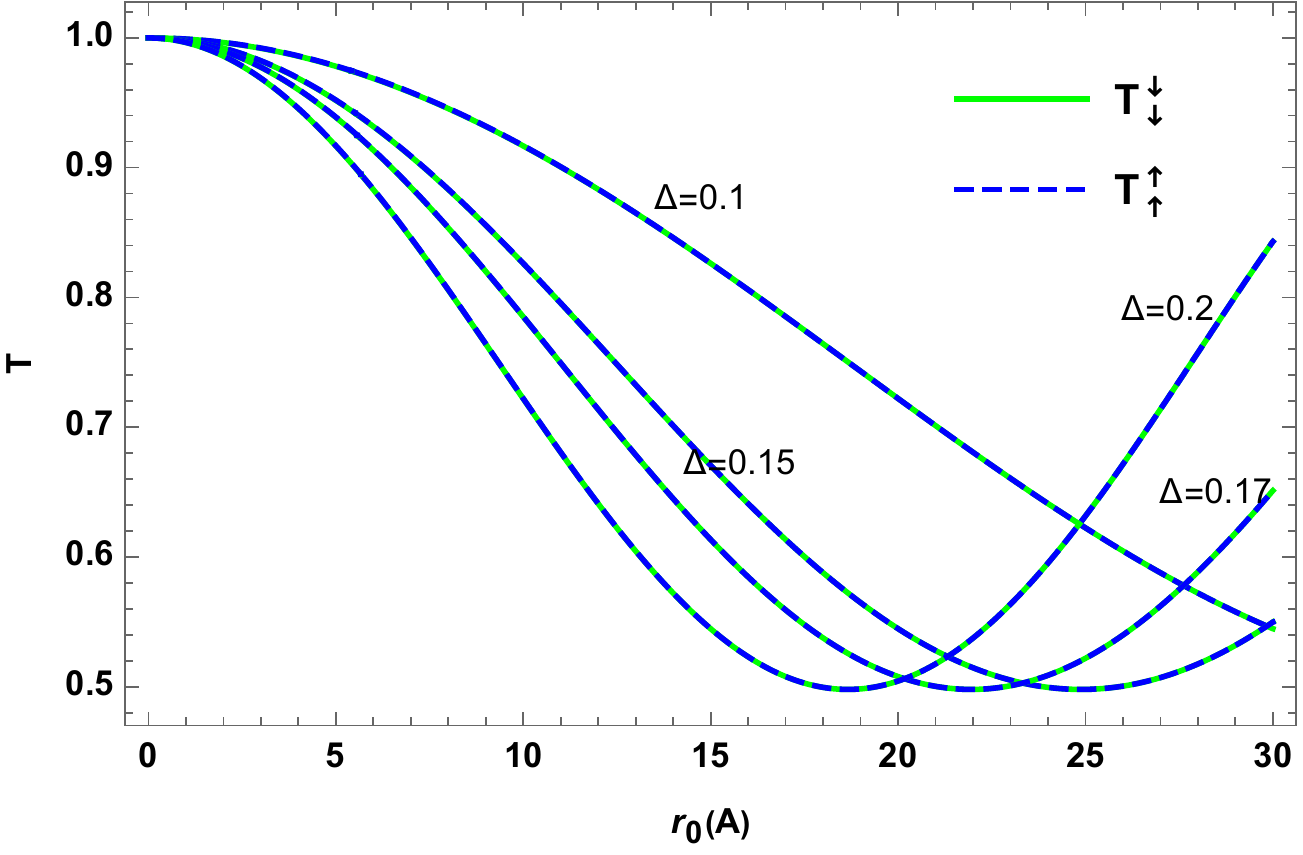}
				\label{subfigdtr}}
			\subfloat[]{
				\includegraphics[scale=0.4]{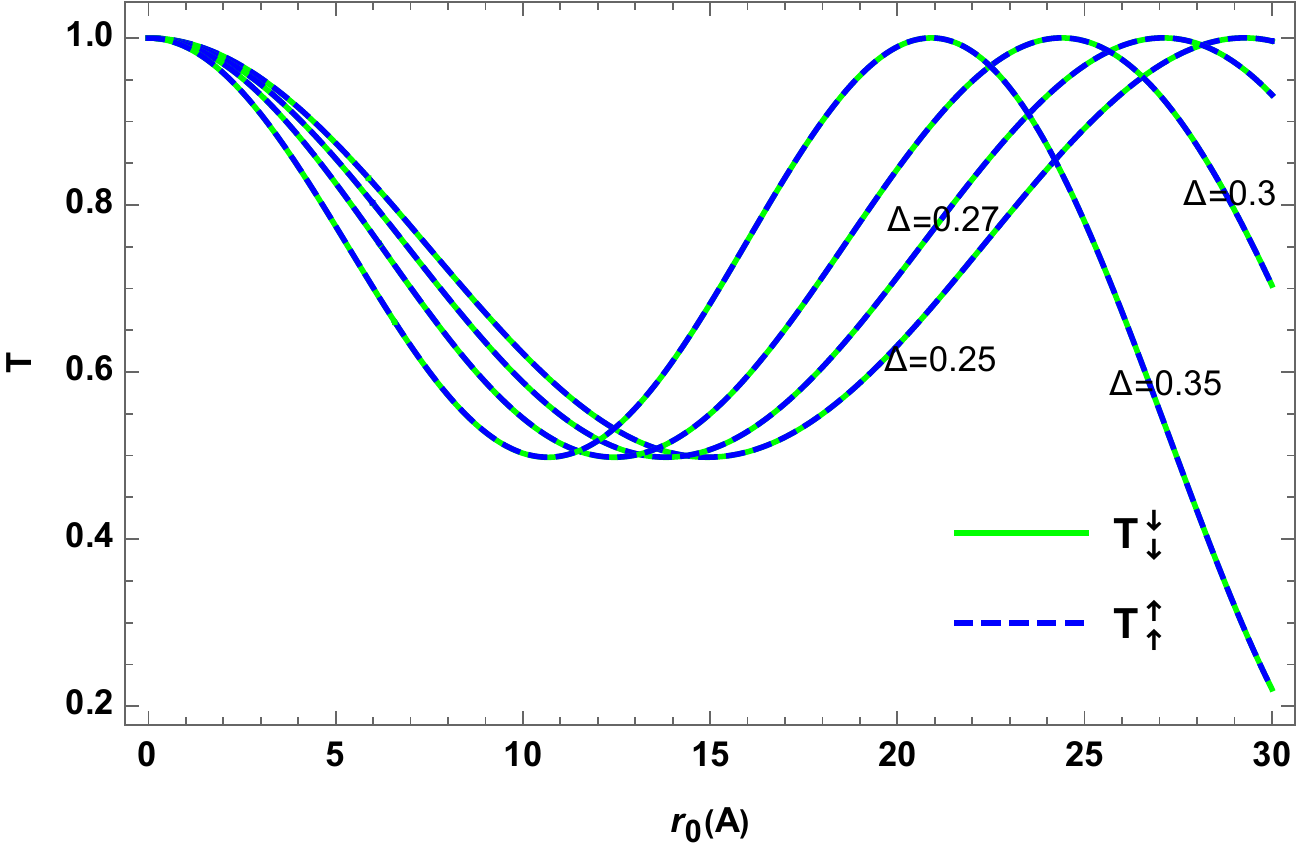}
				\label{subfigetr}}
			\subfloat[]{
				\includegraphics[scale=0.4]{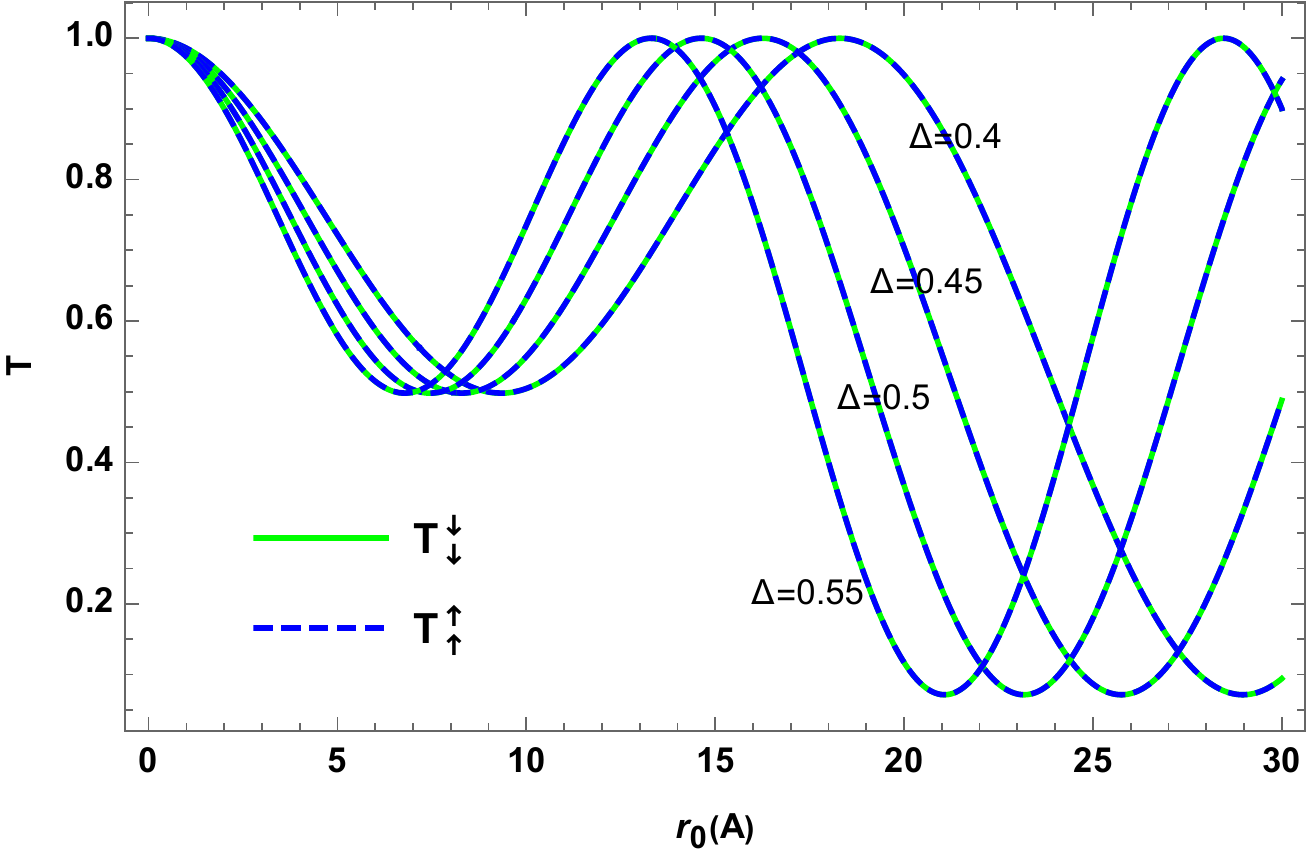}
				\label{subfigftr}}
		\end{center}
		\caption{(color online) The transmission probabilities $T_{\uparrow}^{\uparrow}$ (blue dashed lines) and $T_{\downarrow}^{\downarrow}$ (green solid lines) at normal incidence as a function of the radius $r_{0}$ of ripple  for 
			different values of the band gap $\Delta$  with $E=0.6$ eV
			and 	$\phi=\frac{4\pi}{5}$.}
		\label{delta9}              
	\end{figure}
	
	Figure \ref{delta9}  presents the transmission probabilities with the same spin versus
	the ripple radius for different values of the band gap $\Delta$ 
	with incident energy  $E=0.6$ eV
	and angle	$\phi=\frac{4\pi}{5}$. Our results also show a degradation of the capacity of the spin filter due to $\Delta$. In Figure~\ref{subfigatr} with $\Delta$ in the range $[0,2\ 10^{-3}]$, we observe that  $T_{\uparrow}^{\uparrow}$ (blue dashed lines)
	behaves differently compared to $T_{\downarrow}^{\downarrow}$ (green solid lines)
	when $r_{0}\in [0,16]$, but they approach to each other beyond.
	In Figure~\ref{subfigbtr}, 
	$T_{\uparrow}^{\uparrow}$ and $T_{\downarrow}^{\downarrow}$ mostly coincide
	and then after a critical value of $r_0$ they present the same behavior.
	%
	In Figure~\ref{subfigctr} there is a perfect coincidence between 
	$T_{\uparrow}^{\uparrow}$ and $T_{\downarrow}^{\downarrow}$.
	We observe some oscillations stared to appear from 
	Figure~\ref{subfigdtr} and become clear in Figure~\ref{subfigetr}.
	From $\Delta = 0.55$, the behavior stabilizes and keeps the same minimum and maximum for any value of {$\Delta$ as depicted} in  Figure~\ref{subfigftr}.
	Note that, 
	in all case presented {here} we notice that both of transmissions decrease as long as $\Delta$ increases.
	

	\begin{figure}[ht]
		\begin{center}
			\subfloat[]{
				\includegraphics[scale=0.55]{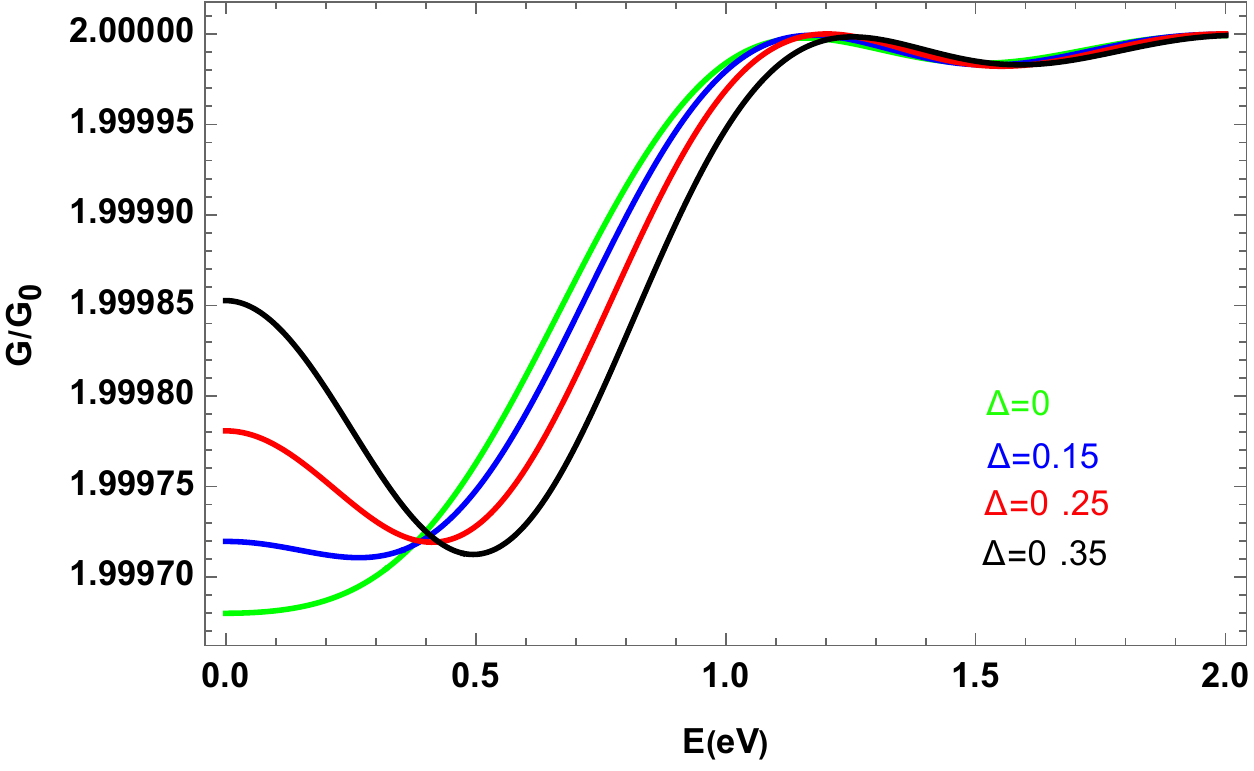}
				\label{subfig:A}}\ \ \ \
			\subfloat[]{
				\includegraphics[scale=0.55]{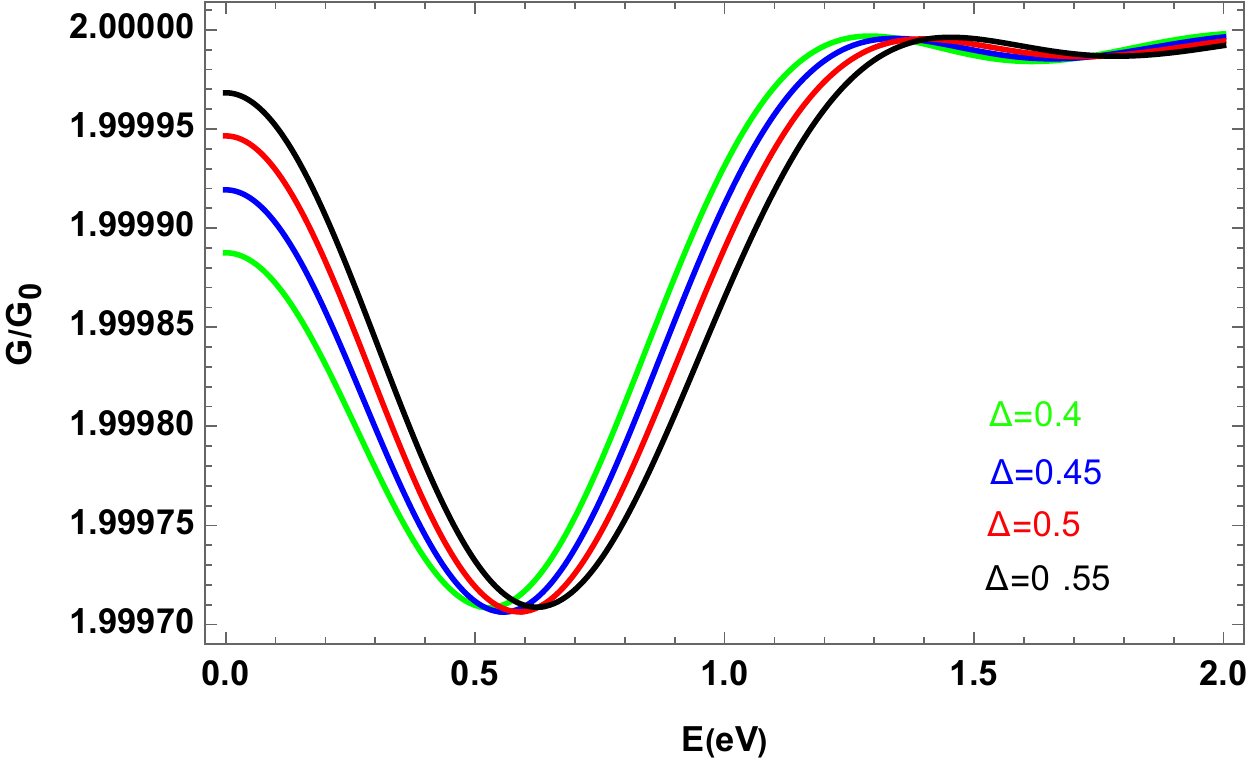}
				\label{subfig:B}}
		\end{center}
		\caption{(color online) The conductance as a function of the incident energies $E$
			for  different  values of the band gap $\Delta$
			with  
			$\phi=\frac{4\pi}{5}$ and  $r_{0}=10\ \angstrom$.}
		\label{delta5}              
	\end{figure}

		We show the conductance as a function of the incident energy in Figure~\ref{delta5} for different values of the band gap $\Delta$ with $r_{0}=10\ \angstrom$ and $\phi=4\pi/5$. Our results show that the conductance changes its behavior on three different zones. According to Figure~\ref{subfig:A}, we can analyze the conductance behavior by considering three zones. Indeed, for the first zone $E\in [0,\sqrt{0.1+\Delta^2}]$ the conductance increases for $\Delta=0$, but
		it 
		starts to decrease by increasing  $\Delta$. Note that at $E = 0$, there is a large spacing for each value of $\Delta$. For the second zone $E\in [\sqrt{0.1+\Delta^2}, 1.2]$ there are rapid increases in conductance but $\Delta$ acts by accelerating. In the third zone where $E$ is beyond the value $1.2$, the conductance becomes insensitive to any increase in energy $E$ and $\Delta$. This behavior is mainly changed in Figure~\ref{subfig:B} under the increase of $\Delta$ because in the zone $E\in [0, \sqrt{0.1+\Delta^2}]$ there is a rapid decrease in conductance for any $\Delta$ and reaches its maximum. In the second zone $E\in [\sqrt{0.1+\Delta^2}, 1.4]$ the conductance increases rapidly in order to stabilize beyond the value $E = 1.4$. Finally, we notice that the increase in $\Delta$ causes some oscillations in the first and third zones.
	\begin{figure}[ht]
		\begin{center}
			\subfloat[]{
				\includegraphics[scale=0.53]{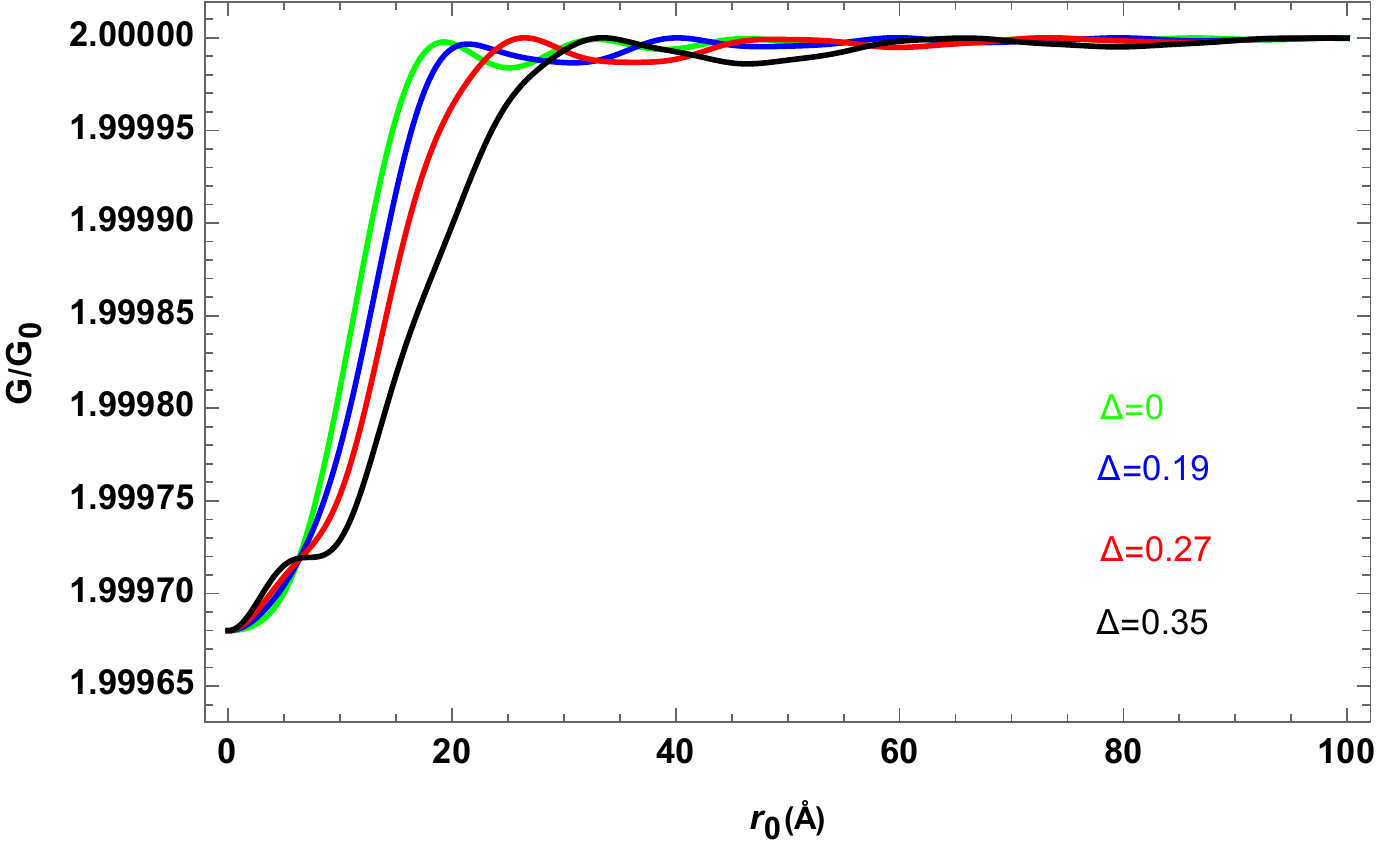}
				\label{subfigacr}}
			\subfloat[]{
				\includegraphics[scale=0.53]{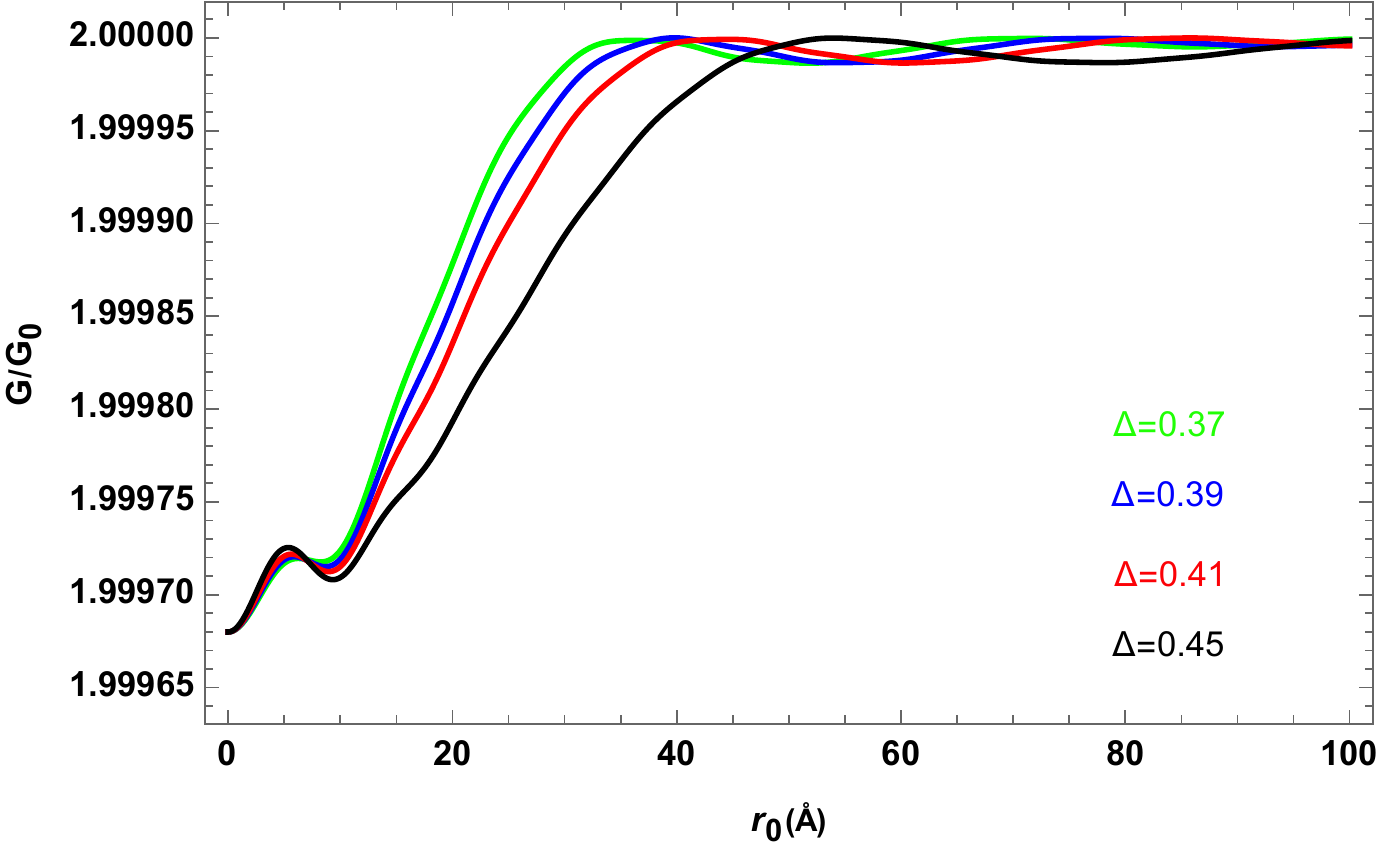}
				\label{subfigbcr}}\\		
		\end{center}
		\caption{(color online) The conductance as a function of the radius of ripple $r_{0}$ for different  values of the band gap $\Delta$
			with   
			$\phi=\frac{4\pi}{5}$ and $E=0.6$ 
			eV.}
		\label{delta11}              
	\end{figure}
	
	We show the conductance as a function of the ripple  radius in
	Figure \ref{delta11} for different  values of the band gap $\Delta$
	with   
	$\phi=\frac{4\pi}{5}$ and $E=0.6$ 
	eV. Our results show that the conductance changes its behavior by increasing $\Delta$. According to Figure~\ref{subfigacr}, one can analyze
	the conductance behavior by considering  three different zones.
	Indeed, for the first zone $r_{0}\in [0,6]$ we have mostly the same increase of conductance what ever the value taken by $\Delta$. For the second zone $r_{0}\in [6,30]$ there are rapid increases in conductance  but
	$\Delta$ acts on by speeding down. In the third zone where $r_0$ beyond the value 30, the
	conductance becomes insensible to any increase of radius and band gap.
	This behavior mostly changed in Figure~\ref{subfigbcr} under the increase of $\Delta$  because in the first zone [0,9] we have a small oscillation
	with an amplitude and increases in the second zone where $r_0$ is in [9, 50]
	afterword it becomes constant in third zone.

	
	\begin{figure}[ht]
		\begin{center}
			\subfloat[]{
				\includegraphics[scale=0.52]{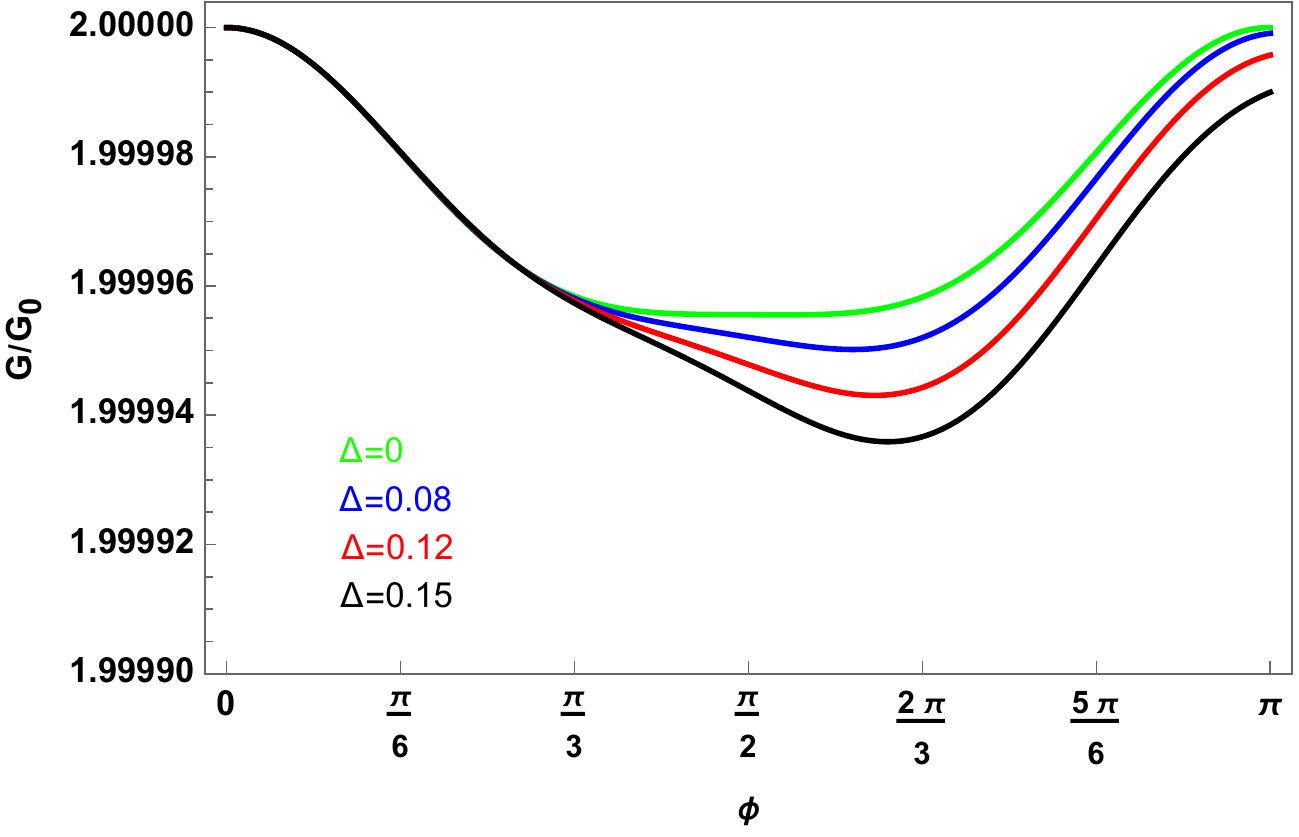}
				\label{subfigac}}
			\subfloat[]{
				\includegraphics[scale=0.52]{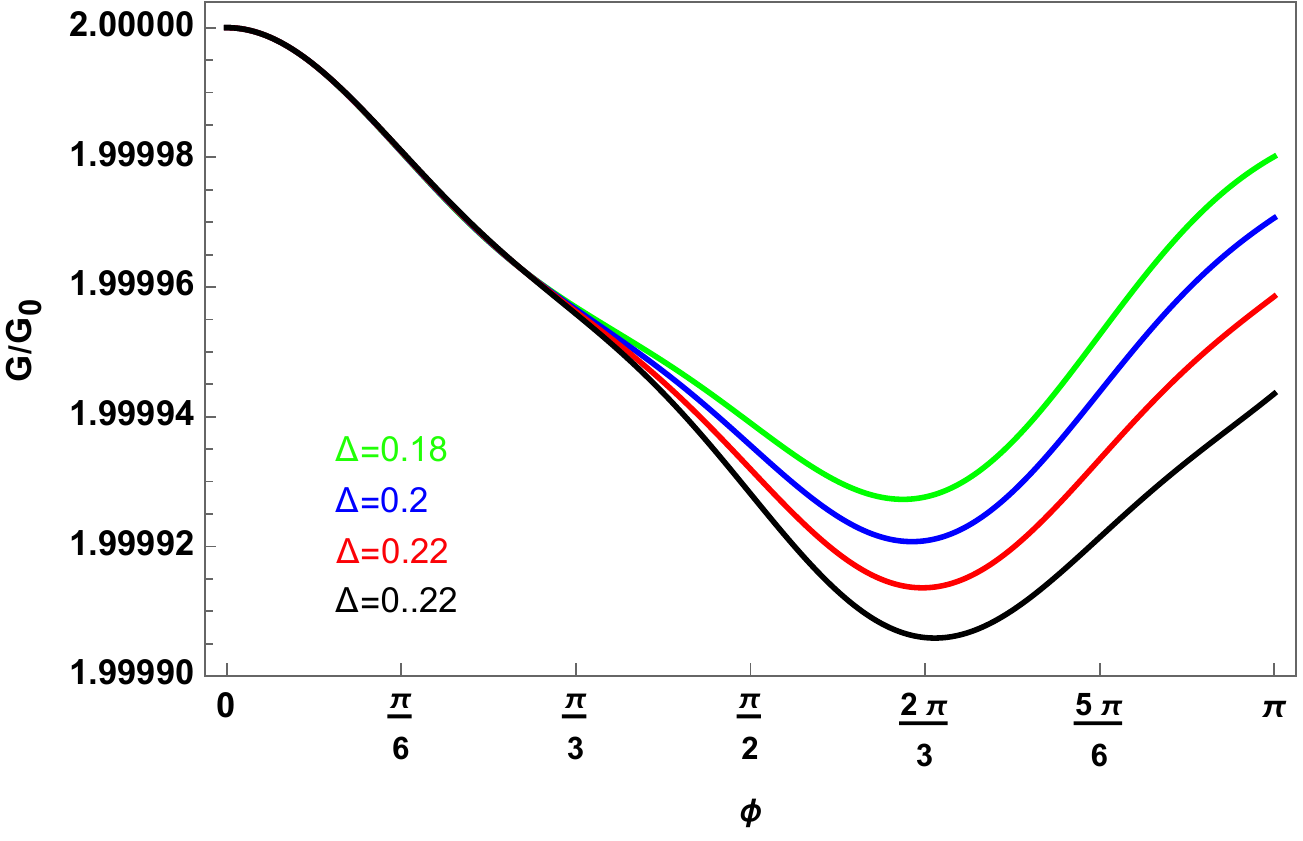}
				\label{subfigbc}}
		\end{center}
		\caption{(color online) The conductance as a function of the ripple  angle  $\phi$  for different  values of the band gap $\Delta$
			$r_{0}=16$ $ \angstrom$ and $E=0.6$ 
			eV.}
		\label{delta10}              
	\end{figure}
	{We show the conductance as a function of ripple angle $\phi$ in Figure~\ref{delta10}
		for different values of the band gap $\Delta$ with $r_{0}=16$ $\angstrom$ and $E=0.6$ eV. Our results show that conductance completely changes its behavior. From Figure~\ref{subfigac}, the behavior of conductance can be analyzed by considering three different areas. Indeed, for the first zone $\phi \in [0,\pi/3]$ the conductance coincides and decreases whatever the value taken by $\Delta$. For the second zone $\phi\in [\pi/3,2\pi/3]$ there are still decreases in conductance but there is a strong widening and reaching the minimum. In the third zone where $\phi$ is beyond the value 2$\pi/3$ the conductance becomes narrow and increases toward a maximum. 
		In Figure~\ref{subfigbc} we observe that the only interesting change is the minimum become deeper compared to those seen in Figure~\ref{subfigac} in addition to a reduction of the  maximum amplitude.}
	
	
	\section{Conclusion}
	We have studied the transport properties  of electrons through the structure of corrugated  graphene in the presence of mass term  at normal incidence ($k_{y}= 0$). By solving the Dirac equation and using the transfer matrix method, the four energy bands are obtained as a function of the opening band gap $\Delta$. Next, we have analyzed the transmissions and reflection channels together with the corresponding conductance. Indeed, we have shown that the presence of band gap $\Delta$ has a visible impact on electron scattering with different initial polarization. 
	
	Furthermore, our  numerical results showed that there is a reflection of electrons with the same spin polarization of the incoming ones {as a manifestation of $\Delta$}. This situation does not exist in {the absence of $\Delta$ as reported in}  \cite{pudlak2015cooperative, pudlak2020spin} and \cite{smotlacha2019spin}. In the case under consideration, there is also the transmission for electrons with the opposite spin polarization, which gradually increase with the deviation of $\Delta$.
	We have also observed the decrease in transmission with same spin polarization. On the other hand, backscattering with the same polarization spin takes place because of the nonzero electron reflection.
	
	{Finally, we mention that the experiment realized by Kuemmeth {\it et al.}  \cite{Kuemmeth2018}  demonstrated that in clean nanotubes
		the spin and orbital motion of electrons are coupled. We think that
		the technique employed in  \cite{Kuemmeth2018}
		can serve as a guide to experimentally reproduce  our work.
		Our results could offer a way to engineer systems towards technological application. Indeed, may be our findings can find important implications for spin-based applications in carbon-based systems, providing a mechanism for all-electrical control of spins \cite{1414} and band gap.}

\end{document}